\documentclass[10pt,emulateapj,apj]{emulateapj}

\shorttitle{Properties of SDSS Galaxies}
\shortauthors{Choi et al.}

\begin{document}
%\twocolumn[
\title{Internal and Collective Properties of Galaxies in the Sloan Digital Sky Survey}

\author{Yun-Young Choi\altaffilmark{1}, Changbom Park\altaffilmark{1}, 
and Michael S. Vogeley\altaffilmark{2}
}
\begin{abstract}

We examine volume-limited samples of galaxies drawn from 
the Sloan Digital Sky Survey to look for relations among internal 
and collective physical parameters of galaxies as faint as 
$M_r=-17.5$. 
The internal physical properties of interest include morphology, 
luminosity, color, color gradient, concentration, size, 
velocity dispersion, equivalent width of $H{\alpha}$ line, 
and axis ratio. Collective properties that we measure include 
the luminosity and velocity dispersion functions.
We morphologically classify galaxies using the three dimensional
parameter space of $u-r$ color, $g-i$ color gradient, and
concentration index.  All relations are inspected separately 
for early and late type galaxies.
At fixed morphology and luminosity, we find that bright ($M_r<-20$)
early type galaxies show very small dispersions in color, color
gradient, concentration, size, and velocity dispersion. These
dispersions increase at fainter magnitudes, where the fraction of 
blue star-forming early types increases. Late types show 
wider dispersions in all physical parameters compared to early 
types at the same luminosity.
Concentration indices of early types are well correlated with 
velocity dispersion, but are insensitive to luminosity and color 
for bright galaxies, in particular. The slope of the Faber-Jackson 
relation ($L \propto {\sigma}^{\gamma}$) continuously changes 
from $\gamma=4.6\pm 0.4$ to $2.7\pm 0.2$ when luminosity 
changes from $M_r = -22$ to $-20$.  
The size of early types is well-correlated with stellar velocity
dispersion, $\sigma$, when $\sigma>100$ km s$^{-1}$.
We find that passive spiral galaxies are well separated from
star-forming late type galaxies at $H{\alpha}$ equivalent width of
about 4.
An interesting finding
is that many physical parameters of galaxies manifest 
different behaviors across the absolute magnitude of
about $M_{\rm \ast} \pm 1$.
The morphology fraction
as a function of luminosity depends less sensitively on
large-scale structure than the luminosity function (LF) does,
and thus seems to be more universal.
The effects of internal extinction in late type galaxies on the 
completeness of volume-limited samples and on the
LF and morphology fraction are found to be very important.
An important improvement of our analyses over most previous works
is that the extinction effects are effectively reduced by excluding
the inclined late type galaxies with axis ratios of $b/a<0.6$.

\end{abstract}
%]
\keywords{galaxies:general -- galaxies:luminosity function, mass function
-- galaxies:formation -- galaxies:fundamental parameters -- galaxies:statistics}
\altaffiltext{1}{Korea Institute for Advanced Study, Dongdaemun-gu, Seoul 130-722, Korea}
\altaffiltext{2}{Department of Physics, Drexel University, 3141 Chestnut Street, 
Philadelphia, PA 19104, USA}

\section{Introduction}

With the large and homogeneous redshift surveys like the Sloan Digital Sky 
Survey (SDSS; York et al.2000) 
and Two Degree Field Galaxy Redshift Survey (2dFGRS; Colless et al. 2001),
our view of the local universe 
became extended out to a few hundred mega parsecs. 
From such dense redshift surveys, the small-scale distribution
of galaxies is also being revealed in more detail.
It is now possible to make very accurate measurements of various relations
among the physical properties of galaxies and the local environment.
This is a difficult task, which should be completed before theoretical modeling 
and interpretations are attempted. The study is complicated because
there are many physical parameters involved and their mutual
relations are non-trivially interrelated.
A further complication is the fact that definition of 
environment depends on the type of
objects chosen to trace the large scale-structure.

One could divide the physical properties of galaxies into internal and
collective ones. Examples of internal physical properties are color,
luminosity, morphology, star formation rate, velocity dispersion,
surface brightness, spectral type, size, and so on. Collective
properties can be the strength and statistical nature of spatial
clustering, peculiar velocity field, luminosity function, velocity
dispersion distribution function, halo mass function, axis ratio
distribution function and so on (see Fig. 1).  
\begin{figure*}
\plotone{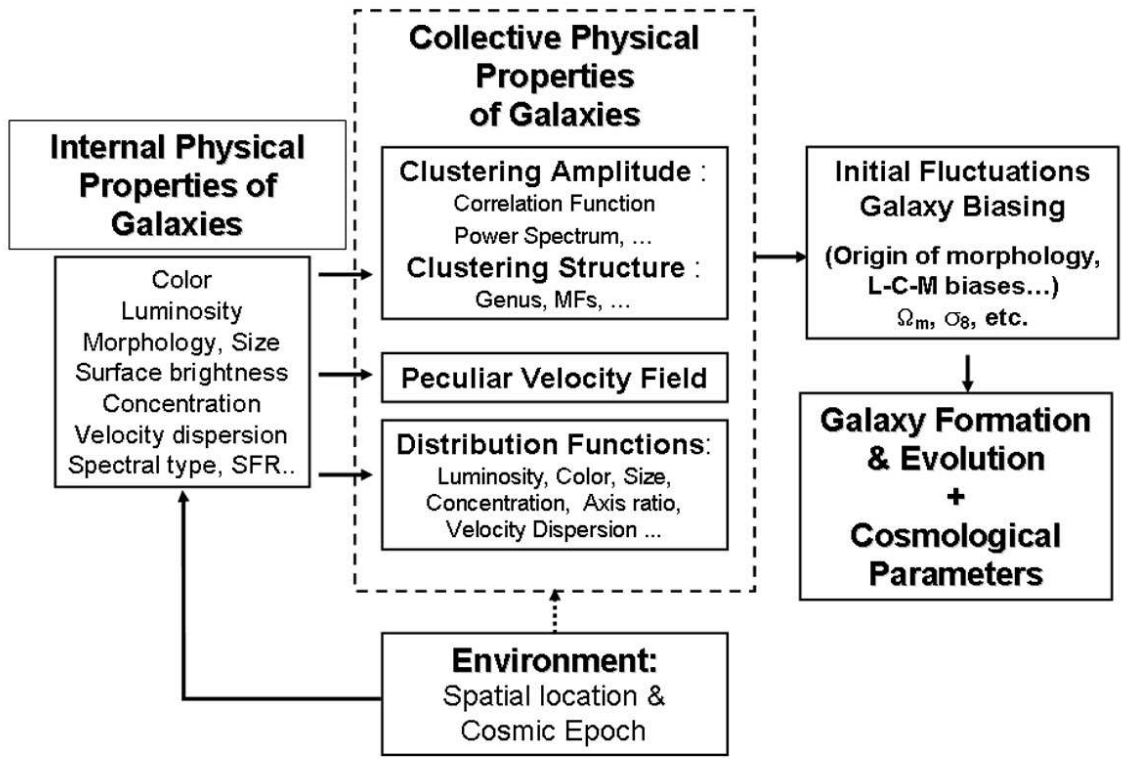}
\caption{Schematic diagram showing some key science for understanding
the structure formation and the background universe.
}
\label{fig1}
\end{figure*}
On the other hand,
the local environments can be spatial or temporal. Dependence on
the latter means the evolution in cosmic time. ''Spatial environment'' can
have various meanings. Traditionally discrete types, such as cluster,
group, field, or void, are used to distinguish among different
environments of galaxies.  Recently, the smoothed number density of
neighboring galaxies was also used to represent the environment. One
could also use other parameters to define the environment of
galaxies. Density gradient, shear fields, and large-scale peculiar
velocity fields can be important parameters affecting the properties of
galaxies. The relations among all these physical observables and
environment can give us information on galaxy formation and the
background universe.
Many authors in the past have investigated the relations 
between environment and galaxy properties (Lewis et al. 2002; 
G\'{o}mez et al. 2003; Goto et al. 2003a; Balogh et al. 2004a, 2004b;
Hogg et al. 2004; Tanaka et al. 2004; Kuehn \& Ryden 2005;
Desroches et al. 2006).

There has been much recent progress in studying the correlation
between the distribution of galaxies and their basic properties,
and the correlation among the properties.
For example, the luminosity function of galaxies subdivided by various
criteria has been measured.
Blanton et al. (2003a) presented the luminosity
functions of the SDSS galaxies defined by galaxy light
profile shape.
Madgwick et al. (2002) measured the luminosity function of 2dFGRS
for different type of galaxies defined by their spectral properties. 
Nakamura et al. (2003) studied the dependence of the luminosity
function on galaxy morphology and the correlation between
galaxy morphology and other photometric properties such as color
and concentration index using a sample of SDSS galaxies with
visually identified morphological types.
Weinmann et al. (2006) performed an extensive study of the dependence
of color, star formation, and morphology of galaxies on halo mass
using galaxy groups selected in the SDSS DR2.
Shen et al. (2003) examined the correlation between 
the size distribution of galaxies and other properties such as
their luminosity, stellar mass and morphological type.
Baldry et al. (2004) analyzed the bivariate distribution of
the SDSS galaxies in the color versus absolute magnitude space.
The fundamental plane and the color-magnitude-velocity dispersion relation
for early type galaxies with SDSS have been investigated
by Bernardi et al. (2003a, 2005)  
Very recently Desroches et al. (2006) showed that the
fundamental plane projections of elliptical galaxies depend
on luminosity using DR4 of the SDSS.
In particular, they found that the radius-luminosity and Faber-Jackson
relations are steeper at high luminosity, which we also found
in our work using a larger SDSS sample with morphologically
cleaner early type samples.
Sheth et al. (2003) and Mitchell et al. (2005) measured
the velocity dispersion distribution function of early type galaxies
carefully selected from the SDSS samples using selection criteria
similar to those of Bernardi et al. (2003a).
Alam \& Ryden (2002) measured the axis ratio distributions for
red and blue galaxies fitted by the de Vaucouleurs and exponential
profiles.

This paper focuses on understanding of
the relationship among many physical properties of galaxies
using a set of volume-limited samples of the SDSS galaxies,
which are further divided into subsamples of constant morphology
and absolute magnitude. This well-controlled experiment
allows us to measure the relations among internal and collective
properties as a function of morphology and luminosity.
In our companion paper we have studied the dependence of various
galaxy properties on the local density environment 
(Park et al. 2006, hereafter Paper II).

%We construct an galaxy catalog from a Large-Scale Structure sample DR4plus 
%(LSS-DR4plus) of New York University Value-Added Galaxy Catalog 
%(NYU-VAGC;  Blanton et al 2005),
%which includes galaxy morphology
%defined by using the accurate morphology classifier of Park et al. (2005)
%and extends the limit of absolute magnitude to be explored 
%to $M_{\rm r} < 17.5$, which corresponds to $M_{\rm r} <M_{\rm *}+2.9$
%(Blanton et al. 2003a; $M_{\rm *}=20.4$). 
%This classifier has been well tested and yields high completeness 
%and reliability reaching 90\%. 
%Using this classifier, we first divide
%the galaxies into Early (E/S0) and Late (S/Irr) morphology types
%to separate the morphology dependency from other possible dependencies
%among the physical properties.
%This morphology catalog is publicly available at 
%http://newton.kias.re.kr/~yychoi/ephoto.html.

This paper is organized as follows. In section 2, we describe 
our catalog and morphology classification and the physical parameters
that we use in Section 3 and 4 to investigate the relations among 
the galaxy properties. We summarize our results in Section 5.
%This paper
%  -- relations among galaxy properties. Both internal and collective
%
%Accompanying paper - environment dependence
%Subsequent papers
%   - (2) other collective properties(clustering amplitude - CF and PS,
%         non-Gaussian properties-genus \& MFs
%   - (3) Evolution,
%   - (4) other tracers (Galaxies and galaxy subclasses, Clusters, LRG, Quasars)
%   - (5) comparisons with galaxy formation mechanisms

\section{Observational Data Set}

\subsection{Sloan Digital Sky Survey}

The SDSS (York et al. 2000; Stoughton et al. 2002;
Adelman-McCarthy et al. 2006) is a survey to explore the large-scale 
distribution of galaxies and quasars by using a dedicated $2.5 {\rm m}$
telescope at Apache Point Observatory (Gunn et al. 2006). The photometric
survey has imaged roughly $\pi$ sr of the northern Galactic cap 
in five photometric 
bandpasses denoted by $u$, $g$, $r$, $i$, and $z$ centered at 
$3551, 4686,6165, 7481,$ and $8931 \AA $, respectively,
by an imaging camera with 54 CCDs (Fukugita et al. 1996; Gunn et al. 1998).
The limiting magnitudes of photometry at a signal-to-noise ratio of $5:1$
are $22.0, 22.2, 22.2, 21.3$, and
$20.5$ in the five bandpasses, respectively. 
The median width of the point-spread function (PSF) 
is typically $1.4\arcsec$, and the photometric
uncertainties are $2 \%$ rms (Abazajian et al. 2004).
%Roughly $5\times10^7 ??$ galaxies are cataloged ???.

After image processing (Lupton et al. 2001; Stoughton et al. 2002; 
Pier et al. 2003) and calibration (Hogg et al. 2001;
Smith et al. 2002; Ivezi\'{c} et al. 2004; Tucker et al. 2006),
targets are selected for spectroscopic follow-up
observation. The spectroscopic survey is planned to continue through 2008
as the Legacy Survey and to yield about $10^6$ galaxy spectra.
The spectra are obtained by two dual fiber-fed CCD spectrographs.
The spectral resolution is $\lambda/\Delta \lambda\sim 1800$, 
and the rms uncertainty in redshift is $\sim 30$ km s$^{-1}$. 
Because of the mechanical constraint of using fibers, 
no two fibers can be placed closer than $55\arcsec$ on 
the same tile. Mainly due to this fiber collision constraint, 
incompleteness of spectroscopy survey reaches about
6\% (Blanton et al. 2003b) 
in such a way that regions with high
surface densities of galaxies become less prominent even after adaptive
overlapping of multiple tiles.
This angular variation of sampling density is accounted for in our analysis.
% I eliminated the abbreviations because they are never used again after this
% paragraph

The SDSS spectroscopy yields three major samples: the main galaxy
sample, the luminous red galaxy sample
(Eisenstein et al. 2001), and the quasar sample (Richards et al.
2002).  The main galaxy sample is a magnitude-limited sample with
an apparent Petrosian $r$-magnitude cut of
$m_{r,\mathrm{lim}}\approx17.77$, which is the limiting magnitude for
spectroscopy (Strauss et al. 2002).  It has a further cut in
Petrosian half-light surface brightness
$\mu_{R50},\mathrm{lim}=24.5$ mag/arcsec$^2$.  
% irrelevant for this paper
%The galaxies in
%the LRGs are selected by color-magnitude cuts in $g$, $r$, and $i$,
%and have spectroscopic magnitude limit of $m_r\sim 19.5$.  
More details about the survey can be found on the SDSS web site.
\footnote{\url{http://www.sdss.org/dr5/}}

In our study of galaxy properties, 
we use a subsample of SDSS galaxies known as the
New York University Value-Added Galaxy Catalog 
(NYU-VAGC;  Blanton et al 2005).
This sample is a subset of the recent SDSS Data Release 5.
One of the products of the NYU-VAGC used here is 
a large-scale structure sample DR4plus (LSS-DR4plus).
%We use the large-scale structure sample DR4plus, which covers *** square degrees
%of the sky, and contains *** galaxies between redshift of $0.001$
%and $0.5$, surveyed as of ***November ***2003.  
For local density estimation in the three-dimensional redshift space
in our accompanying paper we use galaxies within the boundaries 
shown in Figure 1 of Paper II, which improves
the volume-to-surface area ratio of the survey.
%\begin{figure}
%\epsscale{0.8}
%\plotone{fig2.ps}
%\caption{Angular definition of the SDSS sample used for our
%environmental effect study. Solid lines delineate the boundaries of the analysis
%regions in the survey coordinate plane $(\lambda, \eta)$.
%}
%\label{fig2}
%\end{figure}
There are also three stripes in the southern Galactic cap observed by
SDSS. Density estimation is difficult within these narrow stripes, so
for consistency with the samples examined in the accompanying paper,
we do not use them.  The remaining survey region covers
$4464$ deg$^2$.  The primary sample of galaxies used here is a subset
of the LSS-DR4plus sample referred to as ''{\tt void0}'', which is further
selected to have apparent magnitudes in the range $14.5<r<17.6$ and
redshifts in the range $0.001<z<0.5$. These cuts yield a sample of 312,338 galaxies.
The roughly 6\% of targeted galaxies that do not have a measured redshift 
due to fiber collisions are assigned the redshift of their nearest neighbor. 
 
%The large-scale structure
%sample comes with an angular selection function of the survey
%defined in terms of spherical polygons (Hamilton \& Tegmark 2004),
%which takes into account the incompleteness due to mechanical
%spectrograph constraints, bad spectra, or bright foreground stars.
Completeness of the SDSS is poor for bright galaxies with $r<14.5$
because of both the spectroscopic selection criteria (which excludes
objects with large flux within the 3$\arcsec$ fiber aperture; the
cut at $r=14.5$ is an empirical approximation of the completeness
limit caused by that cut) and the difficulty of obtaining correct
photometry for objects with large angular size.  For these reasons,
analysis of SDSS galaxy samples have typically been limited to
$r>14.5$.  This is unfortunate because it limits the range of
luminosity that can be probed; using the magnitude limits of the {\tt
void0} sample the range of absolute magnitude is only about 3.1 at a
given redshift.
To extend the range of magnitude, we attempt to complete the redshift
sample by supplementing the catalog with redshifts of bright galaxies,
first from SDSS itself, and then by matching to earlier redshift
catalogs (the missing galaxies are quite bright and easily observed by
earlier surveys).  We employ another subset of the LSS-DR4plus sample
referred as {\tt bvoid}, which is made before redshift determination
and includes galaxies with $r<14.5$. 
Galaxies in {\tt bvoid} with SDSS redshifts are added to the
catalog. For those objects lacking redshifts, we searched earlier
catalogs.
Out of $5227$ bright objects with $r<14.5$ in the {\tt bvoid} sample
and within our survey region, we assign redshifts to 
3945 galaxies cross matched with the Updated Zwicky Catalog (UZC;
Falco et al. 1999), 807 galaxies from the {\tt bvoid} sample after
redshift determination, 69 galaxies from the Center for Astrophysics
(CfA) redshift
catalog\footnote{\url{http://www.cfa.harvard.edu/\~{}huchra/zcat/zcom.htm}}
(ZCAT 2000 Version), and 4 galaxies from the {\it IRAS} Point Source
Catalog Redshift Survey (Saunders et al. 2000). We found 336 bright objects
to be errors, such as outlying parts of a large galaxies that
were deblended by the automated photometric pipeline, blank fields,
stars, etc. Out of the remaining 66 objects, 30 galaxies are included
in our sample, using redshifts from the NASA Extragalactic Database
(NED\footnote{\url{http://nedwww.ipac.caltech.edu}}). Only
36 galaxies remain in the bright {\tt bvoid} sample with no measured
redshift.  In addition to these bright {\tt bvoid} galaxies, which
have been selected based on their $r$-band magnitudes, we have included
the redshifts of 340 UZC galaxies cross-matched with SDSS objects
in NYU-VAGC, which were not included in the previous step.
Out of 6295 UZC galaxies within our survey region, 242 are still not
matched with SDSS objects in the NYU-VAGC. Ofthe missing UZC
galaxies, 60\% have redshifts smaller than our inner redshift cut of
$z=0.025$, leaving only 99 galaxies missing at larger redshift.

In total, we add 5195 bright ($r<14.5$) galaxies to the {\tt void0}
samples, which yields a final sample of 317533 galaxies.
Volume-limited samples derived from the resulting catalog have nearly
constant comoving number density of galaxies along the radial
direction at redshift $z\ge 0.025$. Thus, we treat our final samples
as having no bright limit at redshift greater than $z=0.025$.

\subsection{Definitions for the Volume-limited Samples}

%To study the effects of environment on galaxy properties it is advantageous 
%for the observational sample to have the lowest possible surface-to-volume ratio 
%as in the topology analysis (Park et al. 2005). For this reason we trim DR4plus 
%as shown in Figure 2, where the gray lines delineate our sample boundaries 
%in the survey coordinate plane ($\lambda, \eta$). 
%Within our sample boundaries, we account for angular variation of the survey 
%completeness by using the angular selection function. We make two arrays of 
%square pixels of size $0.025 \times 0.025$ in the ($\lambda, \eta$) sky 
%coordinates covering our analysis region, and store the angular selection 
%function calculated using the MANGLE routine (Hamilton \& Tegmark 2004). 
%At the location of each pixel, the routine calculates the survey completeness 
%in a spherical polygon formed by the adaptive tiling algorithm (Blanton et al. 2003a) 
%%used for the SDSS spectroscopy. The resulting useful area within the analysis 
%regions with nonzero selection function is 1.362 sr.

In this study we use only volume-limited samples of galaxies selected by
absolute magnitude and redshift limits.  Figure 2 shows
the definitions of our six subsamples in redshift-absolute magnitude space.
\begin{figure*}
\center
\epsscale{0.8}
\plotone{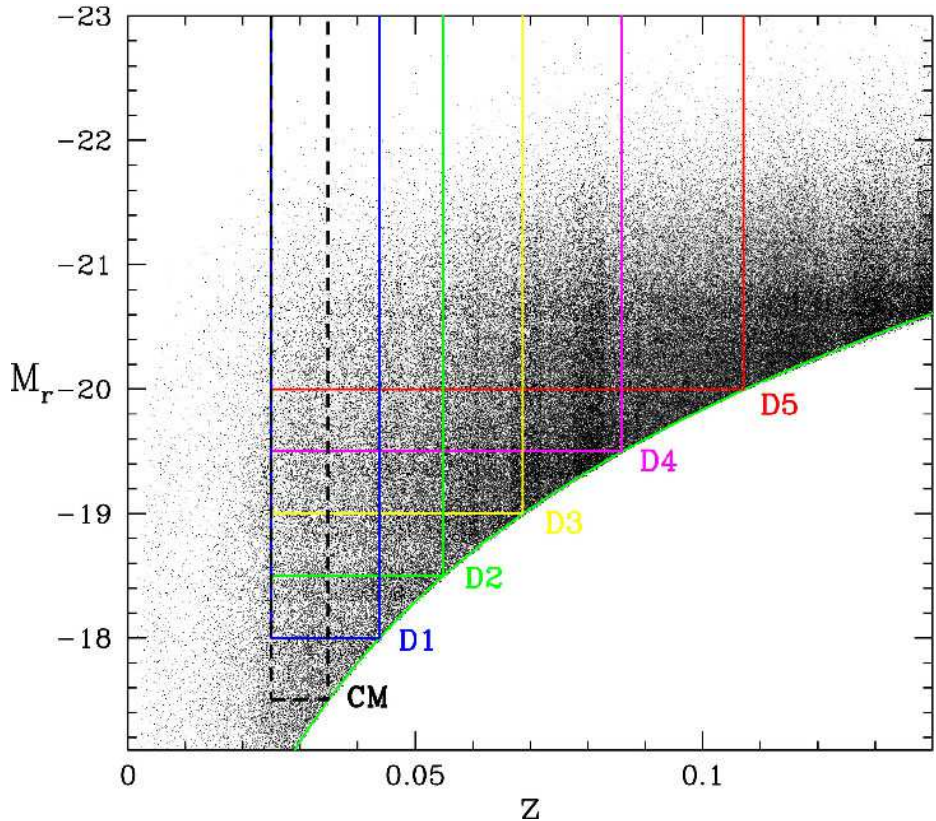}
\caption{Sample definitions of our six volume-limited SDSS samples
in redshift vs. absolute magnitude space.
}
\end{figure*}
The shallowest subsample is labeled CM, which stands for
`color-magnitude'. We employ the CM sample to extend some of our
analyses to fainter absolute magnitude.  The deeper and thicker (along
the line of sight) five samples, from D1 to D5, are the main samples
for study of galaxy properties.
%The LS sample includes the `typical' $L_*$ galaxies ($M_* \approx -20.4$
%for SDSS galaxies, Blanton et al. 2003b), and ranges the entire distance interval
%from $R=74.6$ to 314.0 $h^{-1}$Mpc where our analysis is made. These are the galaxies
%which define the local environment (see below).
The definitions for all samples are summarized in Table 1.

%%%%%%%%%%%%%%%%%%%% Table 1 %%%%%%%%%%%%%%%%%%%%%%%%%
\begin{deluxetable*}{lccccc}
\tabletypesize{\footnotesize}
\tablecolumns{6}
\tablewidth{0pt}
%\begin{center}
\tablecaption{Volume-limited Samples}
\tablehead{
\colhead{Name} &\colhead{Absolute Magnitude} &\colhead{Redshift}&
\colhead{Distance \tablenotemark{a}}&\colhead{Galaxies$(N_{E}\tablenotemark{b})$}&
\colhead{$\bar d$\tablenotemark{c}}}
\startdata
CM& $-17.5>M_{\rm r}$& $0.025<z<0.03484$& $74.6<R<103.7$& 11756 (3467)&3.00\\
D1& $-18.0>M_{\rm r}$& $0.025<z<0.04374$& $74.6<R<129.9$& 20288 (6256)& 3.41\\
D2& $-18.5>M_{\rm r}$& $0.025<z<0.05485$& $74.6<R<162.6$& 32550 (11341)& 3.78\\
D3& $-19.0>M_{\rm r}$& $0.025<z<0.06869$& $74.6<R<203.0$& 49571 (19270)& 4.18\\
D4& $-19.5>M_{\rm r}$& $0.025<z<0.08588$& $74.6<R<252.9$& 74688 (33039)& 4.58\\
D5& $-20.0>M_{\rm r}$& $0.025<z<0.10713$& $74.6<R<314.0$& 80479 (39333)& 5.56\\
\enddata
%\tablecomments{}
\tablenotetext{a}{Comoving distance in units of $h^{-1} \rm Mpc$.}
\tablenotetext{b}{Number of early type galaxies}
\tablenotetext{c}{Mean separation of galaxies in units of $h^{-1} \rm Mpc$}
\end{deluxetable*}
%\end{table}
%%%%%%%%%%%%%%%%%%%%%%%%%%%%%%%%%%%%%%%%%%%%%%%%%%%%%%%%
The comoving distance and
redshift limits of each volume-limited sample defined by an absolute
magnitude limit are obtained by using the formula
\begin{equation}
m_{r,{\rm lim}} - M_{r,{\rm lim}} = 5 {\rm log} (r(1+z)) + 25 + \bar{K}(z) + \bar{E}(z),
\end{equation}
where $\bar{K}(z)$ is the mean $K$-correction, $\bar{E}(z)$ is the mean 
luminosity evolution correction,
and $r$ is the comoving distance corresponding to redshift $z$. 
We adopt a flat $\Lambda$CDM cosmology with density parameters
$\Omega_{\Lambda}=0.73$ and $\Omega_m=0.27$ to convert redshift to 
comoving distance. To determine sample boundaries
we use a polynomial fit to the mean $K$-correction,
\begin{equation}
\bar{K}(z)=3.0084(z-0.1){^2}+1.0543(z-0.1)-2.5 {\rm log} (1+0.1).
\end{equation}
We apply the mean luminosity evolution correction given by
Tegmark et al. (2004), $E(z)=1.6(z-0.1)$. 
The rest-frame absolute magnitudes of individual galaxies are computed
in fixed bandpasses, shifted to $z=0.1$, using Galactic reddening corrections
(Schlegel et al. 1998) and 
$K$-corrections as described by Blanton et al. (2003c).
This means that a galaxy at $z=0.1$ has a $K$-correction of
$-2.5 {\rm log}(1+0.1)$, independent of its spectral energy distribution (SED).
We have applied the same mean luminosity evolution correction formula to 
individual galaxies.
%\begin{equation}
%x = -R \sin\lambda,
%y = R \cos\lambda \cos\eta,
%z = R \cos\lambda \sin\eta.
%\end{equation}
%The $y$-axis is at the center of the survey coordinate and radially outward,
%the $z$-axis is toward the $(\eta, \lambda) = (90^{\circ}, 0^{\circ})$ direction, 
%and the $x$-axis is at the $(0^{\circ}, -90^{\circ})$ direction. 
\subsection{Physical Parameters of Galaxies}
The physical parameters we consider in this study are $^{0.1}(u-r)$
color, absolute Petrosian magnitude in the $r$-band $M_r$, morphology
(see below), Petrosian radius, axis ratio, concentration index, color
gradient in $^{0.1}(g-i)$ color, velocity dispersion, and equivalent
width of the H$\alpha$ line.  

To compute colors, we use extinction and
$K$-corrected model magnitudes. The superscript 0.1 means the
rest-frame magnitude $K$-corrected to the redshift of 0.1. All our
magnitudes and colors follow this convention, and the superscript will
subsequently be dropped.
To measure some of the physical parameters of galaxies, we retrieve
the $g$- and $i$-band atlas images and basic photometric parameters of
all galaxies in our sample from the archive of photometric
reductions conducted by the Princeton/NYU group.
\footnote{\url{http://photo.astro.princeton.edu}}
To take into account flattening or inclination of galaxies,
we use elliptical annuli in all parameter calculations, and
the isophotal position angle and axis ratio in the $i$-band are
used to define the elliptical annuli.
The $g-i$ color gradient is defined to be the color difference between
the region with $R<0.5R_{\rm Pet}$ and the annulus with $0.5R_{\rm
Pet}<R<R_{\rm Pet}$, where $R_{\rm Pet}$ is the Petrosian radius.  
The (inverse) concentration
index is defined by $R_{50}/R_{90}$ where $R_{50}$ and $R_{90}$ are
the semi-major axis lengths of ellipses containing 50\% and 90\% of
the Petrosian flux in the $i$-band image, respectively.  
The $g-i$ color gradient, concentration index, and isophotal axis ratios 
are corrected for the effects of seeing.  
To do so, we first generate a large set of Sersic model
images with various scaling lengths, slopes, and axis ratios convolved
with a range of PSF sizes.  For an observed galaxy image with a given
size of the PSF we look for a best-fit Sersic model when the Sersic
model is convolved with the PSF to find the true Sersic index and the
true axis ratio.  The fit is made at radii between $0.2$ and
$1.0 R_{\rm Pet}$ to avoid the central region, whose profile is strongly
affected by seeing.  When the best-fit Sersic models are found in the
$g$ and $i$ bands, the $g-i$ color gradient and concentration index
are calculated for the true Sersic models and their differences from
those of convolved images are used to correct the color gradient of
the observed image for the seeing effects.

We classify galaxy morphologies into early
(ellipticals and lenticulars) and late (spirals and irregulars) types
using the automated galaxy morphological classification method
 given by Park \& Choi (2005),
who showed that
incorporating the $g-i$ color gradient into the classification
parameter space allows us to successfully separate the two populations
degenerated in $u-r$ color and 
the concentration index helps us separate red disky spirals
from early type galaxies (see Fig. 1 of Park \& Choi 2005).
It should be noted that the ``morphological'' properties of galaxies
can change when different bands are adopted. The difference in 
morphology in different bands is exactly the color distribution,
which is reflected in our color gradient parameter.
To define the classification boundaries in this multi-parameter space,
they used a training set of 1982 SDSS galaxies whose morphological
types are visually identified.
The distributions of the early and late type galaxies in sample D2
are shown in Figures 6$a$ and 6$b$, below.
The reader is referred to Park \& Choi (2005) for further details
of the methods and various tests. 
We have visually examined color images of a set of faint galaxies retrieved
by the SDSS Image List, and confirmed that the resulting morphological
classification is highly successful
with completeness and reliability reaching 90\%.  For the
volume-limited samples D2 and CM we made an additional visual
check to correct possible misclassifications of the automated scheme for
blue early types (those below the straight line in Fig. 3$a$) and red
late types (those redder than $2.4$ in $u-r$ color). The morphological
types of 1.9\% of galaxies, which are often blended or merged objects,
were changed by this procedure. 
This morphology classification allows us to make very accurate 
measurements in this study compared with previous work. 
We discuss this further in section 3.2.
%-- [end]

\section{Relations Among Physical Parameters of Galaxies}

We find that absolute magnitude and morphology are the most important
parameters characterizing physical properties of galaxies, in the sense
that other parameters show relatively little scatter once absolute
magnitude and morphology are fixed.  This is particularly true for
early type galaxies.  We begin this section by showing the dependence
of various physical parameters of galaxies on their absolute
magnitudes. In each case, we separately examine samples of early
(E/S0) and late (S/Irr) morphological types.

\subsection{Variation with Absolute Magnitude}

Figure 3$a$ shows the color-absolute magnitude diagram of galaxies in
the CM and D2 samples. 
\begin{figure*}
%\hbox{\hspace{2.3in} E/S0 \hspace{1.7in} S/Irr}
\epsscale{1}
\plotone{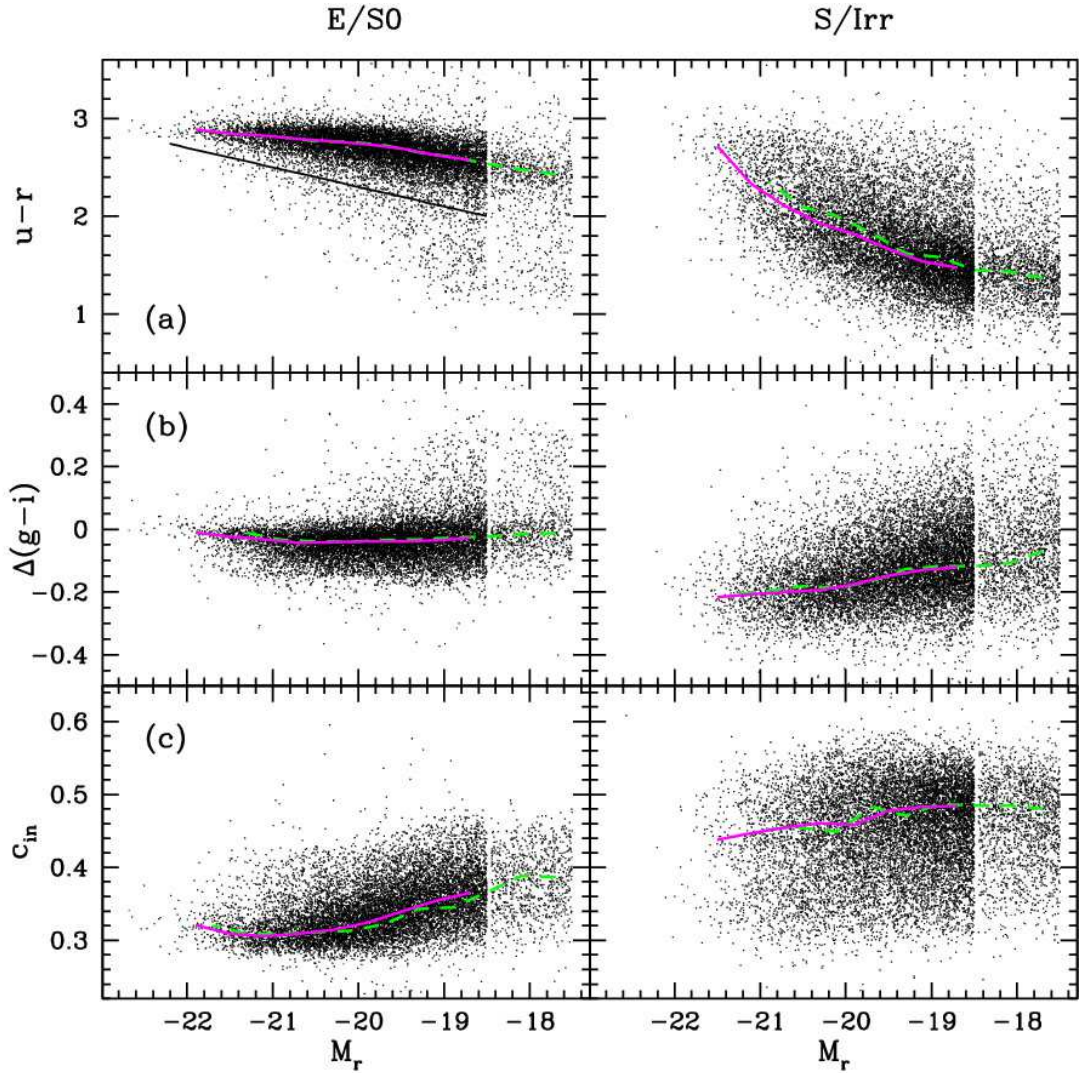}
\caption{Relations between the absolute magnitude and ($a$) $u-r$ color,
($b$) $\Delta(g-i)$ color gradient, and (c) $c_{\rm in}$ (inverse)
concentration index of galaxies in the D2 (brighter than
$M_r = -18.5$) and CM (fainter than $-18.45$) volume-limited
subsamples of the SDSS. E/S0 morphological types are shown
in the left panels, and S/Irr types are shown in the right panels.
Curves delineate the locations of the most probable parameter values
as a function of absolute magnitude. The solid curve is for the D2 sample,
and the dashed curve is for the CM sample.
}
\label{fig3}
\end{figure*}
Galaxies in the CM sample are shown only for
$-17.5 > M_r > -18.45$. The left panel shows the distribution of early
types, and the right panel shows that of late types.  The solid and
dashed lines delineate the most probable $u-r$ color of galaxies in
the D2 and CM samples, respectively.  (Here and throughout, the ``most
probable'' parameter value within
0.4 absolute magnitude bin is the mode of the distribution of that parameter.)  
The most probable color of the
red sequence galaxies has a break at $M_r \approx -19.6$ and has
slopes of about $-0.07$ at the bright side and about $-0.15$ at the
faint side.  We plot only those late type galaxies with axis ratios
$b/a > 0.6$, to reduce the biases in absolute magnitude, color, color
gradient, etc., due to internal extinction.  Colors of late types have
a much wider dispersion. Their most probable color rapidly becomes
bluer as galaxies become fainter, but is very insensitive to absolute
magnitude fainter than $M_r=-18$, approaching $u-r\approx 1.1$ at the
very faint end ($M_r \sim -14$, not shown).

Baldry et al. (2004) reported a slope of about $-0.08$ for
luminous red galaxies selected from the SDSS,
which is consistent with our measurement.
One significant difference between Baldry et al.'s and our
result is the color-magnitude relation of the late type (blue
distribution) galaxies at the bright side.
In their results the slope of the relation becomes less steep
long before the color-magnitude relation for late types touches
that of early types as one, toward brightest magnitude.
On the other hand, our result shows no such bending until it
approaches the relation for early types.
The difference is probably caused by extinction of late type galaxies.

Figure 3$a$ shows that the color distribution of late type galaxies
overlaps with early types at all magnitudes, and vice versa.  The solid
straight line in Figure 3$a$ roughly divides the sequences of red and
blue galaxies. However,
about 24\% of spiral/irregular galaxies brighter than
$M_r=-18.5$ are above this line when all
inclinations are allowed.
Among all galaxies above this line, 32\% are actually
late types. Therefore, an early type sample generated by a simple cut
in the absolute magnitude versus color space must take into account
this huge contamination. We also find a significant
fraction of early types located outside the main red sequence at
faint magnitudes.  At absolute magnitudes between $-18.5$ and $-19.0$,
about 10\% of early types are located below the straight line shown
in Figure 3$a$. This outlying fraction is larger at fainter magnitudes. Some of
these blue early type galaxies may be actually spirals whose disks
look too faint for both our automated and visual classification
schemes to classify correctly, but the trend seems robust.  

Again, it should be noted that even though the bimodality in galaxy
color is mainly due to the bimodality in galaxy morphology, there is a
significant overlap in color space 
between the two morphological classes.
With few exceptions, very blue ($u-r<1.8$) early type galaxies
show strong emission lines in their spectra, indicating active star
formation (see Fig. 6$c$).
%Since their surface brightness 
%distribution is smooth and colors are uniform, 
%the star formation is global and homogeneous within them.
The distributions of colors of early and late types at a given
absolute magnitude are not Gaussian. 
The colors of early types are
skewed to blue color, and those of late types are skewed to red color.
%A Gaussian distribution is a reasonable approximation only for early
%type galaxies much brighter than $M_*$, which is $-20.16\pm 0.04$ in
%the case of the D2 sample.
The blue sequence appears less prominent when the data is contaminated
by internal extinction effects (see Fig. 12, below). 

Figure 3$b$ shows distributions of galaxies in color gradient, $\Delta(g-i)$, 
versus absolute magnitude, $M_r$.
We find that the majority of E/S0's ({\it left panel}) have a weak negative
color gradient; the envelope is slightly bluer than the core.
Their most probable color gradient is surprisingly constant at
$\Delta(g-i) = -0.035\pm 0.007$ at all magnitudes explored, with
a weak trend for more positive gradient toward faint and bright
magnitudes. Interestingly, the E/S0's with $M_r\approx M_\ast-0.5$ have
the minimum color gradient. It should be noted that the color gradient
of the brightest early types approaches zero, meaning that the
brightest early types are reddest and most spatially homogeneous in color. 
Although the most probable color gradient 
has a nearly constant of absolute magnitude, the number of E/S0's
with blue cores increases strongly at absolute magnitudes fainter than 
$M_*$. 

The color gradient of late type galaxies is more negative (i.e., bluer
envelope) than that of early types and is a slowly increasing 
function of absolute magnitude 
Faint late types are more gas rich and have more uniform
internal star formation activity, while bright late types have red
central bulges and blue star forming disk structures.

Figure 3$c$ shows that the (inverse) concentration index $c_{\rm in}$ of the
early types ({\it left panel}) is nearly independent of absolute magnitude
when $M_r < M_*$, but rises rather steeply at fainter magnitudes.  The
most probable concentration index of the bright early types is close to
that of a de Vaucouleurs profile ($c_{\rm in}=0.29$), even though it
slightly increases at the brightest end. The brightest early types
might be less concentrated because of more recent mergers, and the
galaxies only slightly brighter than $M_\ast$, with $M_r \sim M_{\ast}-0.9$
might be dynamically older (more relaxed).
For late type galaxies the most probable value is close to the value
an exponential disk ($c_{\rm in}=0.44$) at the brightest end, albeit with
a large dispersion.

The Petrosian radius in units of $h^{-1}$ kpc
as a function of absolute magnitude is shown in Figure 4$a$.  
The Petrosian radius is calculated by using the elliptical annuli,
and is typically larger than the value in the SDSS Photometric
database which uses the circular annuli.
Early type galaxies ({\it left panel}) exhibit a tight
correlation between physical size and luminosity, and
the relation has a significant curvature.
The slope is very steep 
($d{\rm log}R_{\rm Pet}/dM_r \approx -0.28$ at $M_r \approx -21.5$)
at the brightest end.
The late types have larger $R_{\rm Pet}$, and their size relation with
absolute magnitude is less curved.
The curvature of the
relation for both early and late types means that fainter galaxies
have lower surface brightness. As a guide the radius-magnitude
relation in the case of constant surface brightness ($L \propto \Sigma
R^2$ where $L$ is luminosity and $\Sigma$ is surface brightness) is
drawn in both panels. 
The slope of this relation to the bright end
of the late types tells that the brightest spiral galaxies, with $M_r
\le -21$, have surface brightness approximately independent of
luminosity. This relation is also true for the early types
in the regime $-21 \le M_r \le -20$ fainter than for the late types.  
Desroches et al. (2006) also found that the radius-luminosity relation
of elliptical galaxies is steeper at high luminosity. 
Shen et al. (2003) measured the size distributions of early and late
type galaxies, divided by the concentration index or the Sersic index,
and found that they are well fitted by a log normal distribution.
\begin{figure*}
%\hbox{\hspace{2.3in} E/S0 \hspace{1.7in} S/Irr}
%\epsscale{0.8}
\plotone{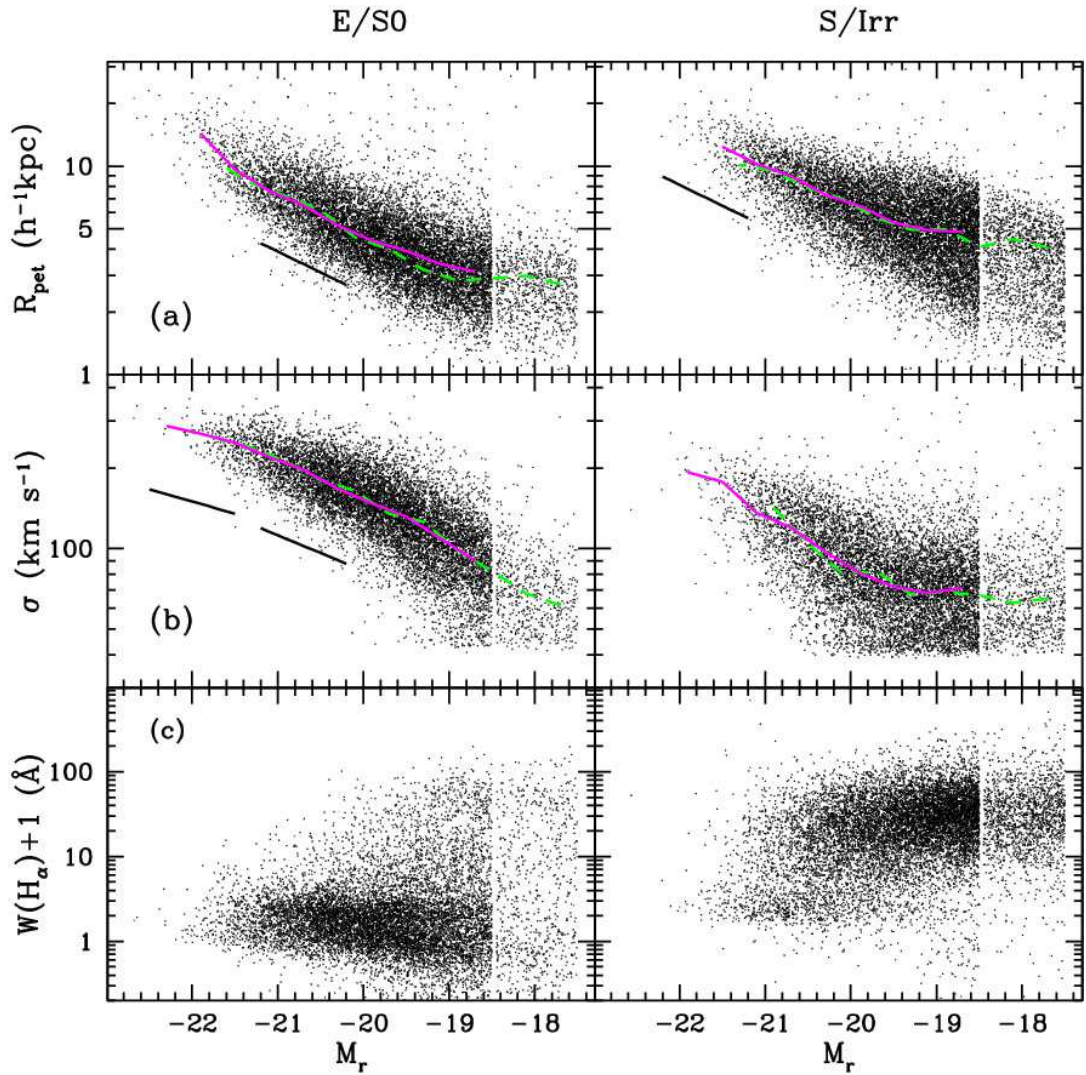}
\caption{
Relations between the absolute magnitude and ($a$) the Petrosian
radius, ($b$) velocity dispersion, and ($c$) equivalent width of the 
$H{\alpha}$ line of galaxies in the D2 (brighter than
$M_r = -18.5$) and CM (fainter than $-18.45$) volume-limited
subsamples of the SDSS survey. E/S0 morphological types are shown 
in the left panels, and S/Irr types are shown in the right panels.
Curves have the same meanings as in Fig. $3a-3c$.}
\end{figure*}

Figure 4$b$ shows the variation with absolute magnitude of velocity
dispersion, as measured by an automated spectroscopic pipeline called {\tt
IDLSPEC2D} version 5 (D. Schlegel et al. 2007, in preparation).  
Galaxy spectra
of the SDSS are obtained by optical fibers with radius of $R_{\rm
fib}=1.5\arcsec$. The finite size of this aperture smooths the
central velocity dispersion profile. To correct the central velocity
dispersion for this smoothing effects we adopt a simple aperture
correction, $\sigma_{\rm corr} = \sigma_{\rm fib} (8R_{\rm
fib}/R_{0})^{0.04}$ (J{\o}rgensen et al. 1995; 
Bernardi et al. 2003b), where $R_0$ is the equivalent circular 
effective radius $(b/a)^{1/2}_{\rm deV}r_{\rm deV}$ 
where $r_{\rm deV}$ is the seeing-corrected effective angular 
radius along major axis of the galaxy
from model fits to a de Vaucouleurs profile in the $i$ band.  
Because the measured velocity
dispersion is systematically affected by the finite resolution of the
spectrographs at small values, the calculation of the most probable
velocity dispersion is restricted to galaxies with $\sigma>40$ km s$^{-1}$.
The errors in the measured velocity dispersions are about $\sim 15\%$ at
70 km s$^{-1}$ and $\sim 30\%$ at 50 km s$^{-1}$. 
The dispersion in $\sigma$ is typically $\Delta{\rm log}\sigma=0.11$,
and the measurement error is typically 11 km s$^{-1}$ (see section 4.2).
From these data we fit the slope, $\gamma$, defined by $L \approx
\sigma^\gamma$. At the bright end, near absolute magnitude $M_r=-22$,
we find $\gamma=4.6\pm 0.4$, while the slope at magnitudes between
$M_r = -21$ and $-20$ drops to $\gamma=2.7\pm 0.2$ (All five
volume-limited samples are used for the measurements).  For
comparison, we plot two straight lines with slopes of $\gamma=4$
({\it upper left}, Fig. 4$b$) and 3 ({\it lower right}, Fig. 4$b$), 
the former being the slope of the
Faber-Jackson relation.
Recently, Desroches et al. (2006) also found that the slope
$\gamma$ is steeper at high luminosity, which is consistent with our
findings.
The measured velocity dispersion of spiral galaxies brighter than
$M_r=-20.5$ is $70\% \pm 2$\% of that of early types (or $\sigma_{\rm
late} \approx \sigma_{\rm early}/\sqrt{2}$) at a given absolute
magnitude.
%This is consistent with the fact that the velocity dispersions obtained from
%the SDSS fiber spectra are given by the random motions of stars 
%corresponding to virial mass in the case of early type galaxies but
%are given by the circular motion of disk stars balanced with the centrifugal
%force in the case of spiral galaxies.

Because SDSS fiber spectra sample the light from only the central $1.5\arcsec$
radius region of galaxies, the measured
spectral parameters are not fully representative of the physical
properties of spiral galaxies, 
which tend to have large variation of star formation activity from the
bulge to outer disk.  Nevertheless, parameters based on fiber spectra,
such as line widths and those from the principal component analysis,
are often used to characterize the star formation activity,
morphological types, etc. (e.g. Tanaka et al. 2004; Balogh et
al. 2004b). Thus, for comparison, we show such results for our
samples.

The right panel of Figure 4$c$ shows the equivalent width of the
$H\alpha$ line of late types as a function of $M_r$.
The equivalent width of the $H{\alpha}$ line
is often used as a measure of recent star formation
activity (or nuclear activity).
The majority of late types have large $W(H\alpha)$. The plot
shows that fainter galaxies tend to have larger $W(H\alpha)$. This
result should be considered carefully because the fiber spectra
systematically miss the light from the outer disks of bright large
galaxies, which might be active in star formation. The figure also
shows that there is a class of late types that 
have weak $H\alpha$ line emission. The placement in this diagram of
some of these objects may be due to the finite fiber size, but some of
them are genuine passive spirals (Couch et al. 1998; 
Poggianti et al. 1999; Goto et al. 2003b; Yamauchi \& Goto 2004).

The left panel of Figure 4$c$ shows $W(H\alpha)$ 
of early type galaxies.  We find
that the number of early types with central star formation activity
increases sharply at absolute magnitudes fainter than $\sim M_\ast$. 
We find similar trends for $u-r$ color and $\Delta (g-i)$
color gradient (see also Fig. 6, below).  Early types with central star
formation activities have been studied and some of them are
$\rm{E+A}$ galaxies (Goto et al. 2003c; Fukugita et al. 2004; 
Quintero et al. 2004; Goto 2005; Park \& Choi 2005; Yamauchi \& Goto 2005).

Figure 5 shows the isophotal axis ratio of early ({\it left panel}) and
late ({\it right panel}) types as a function of absolute magnitude. 
\begin{figure*}
%\hbox{\hspace{2.3in} E/S0 \hspace{1.7in} S/Irr}
%\epsscale{0.8}
\plotone{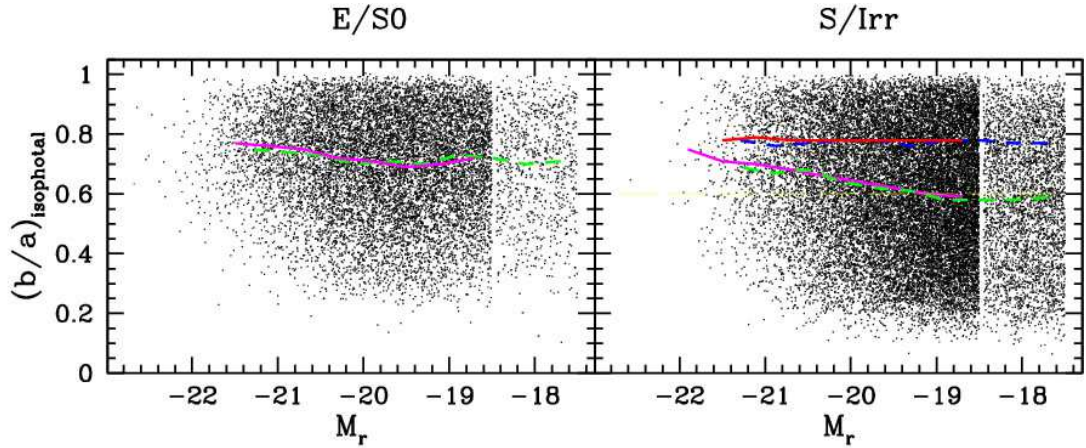}
\caption{
Isophotal axis ratios of the early ({\it left}) and late ({\it right}) 
type galaxies in sample D2 ({\it solid magenta line}) and 
CM ({\it dashed green line}).
Curves are the median values. For late types we added the median curves, 
in red and blue colors, derived from only those with $b/a > 0.6$. 
}
\end{figure*}
We choose the
isophotal axis-ratio rather than axis ratios from de Vaucouleurs or
exponential profile fits 
that are computed by the automated SDSS pipeline
in our study because the isophotal position angles
correspond most accurately to the true orientation of the major axis.
The exponential profile fit is often bad particularly for barred
spiral galaxies.
Curves in Figure 5 are the median axis ratios of galaxies in the D2
({\it solid line}) and CM ({\it dashed line}) samples. 
We also draw the median curves in red and blue colors
by using {\it only} galaxies with $b/a >0.6$.

We clearly see that
early type galaxies tend to be rounder at brighter magnitudes.
Figure 5 demonstrates that the absolute magnitudes of
inclined late types with $b/a < 0.6$ seriously suffer from
dimming due to the internal absorption,
while the effects are ignorable for those
with $b/a > 0.6$. 
See also Figure 12, below and discussion in section 4.3. 

\subsection{Variation with Color}
Figure 6$a$ shows the distribution of E/S0 ({\it left panel}) and S/Irr 
({\it right panel}) galaxies in the D2 sample in the $u-r$ color 
versus $\Delta(g-i)$ color gradient space.
Our morphological classification is based primarily on this parameter space.
Note that the sharp boundaries shown in Figure 6a are due to
a choice of our morphological criteria in the $u-r$ versus $g-i$ color
gradient space, which is determined to correlate
well with visually classified morphologies.
Scatter across the classification boundaries is caused by the
concentration index constraint 
on early types (see Park \& Choi 2005
for details) and corrections made by visual inspection.
Most early type galaxies fall within a strong concentration near
$u-r=2.8$ and $\Delta(g-i)=-0.04$ in this plane. Because the color
of the red sequence becomes bluer at fainter magnitudes (see Figure 3a),
the center of this concentration moves to bluer $u-r$ when one examines
fainter early types. There is also a trail of early types towards bluer
colors and more positive color gradients. The classification boundary
separating this blue early types from late types has been determined
from a training set that showed such a trend (Park \& Choi 2005). 
The majority of galaxies in this
early type trail are fainter than $M_\ast$ ({\it blue points}), 
%--- [this part is added]
often show emission lines (see early types in Fig. 6$c$), and 
live mainly at intermediate and low density environments (see
early types in Fig. 11$a$ of Paper II). 
These blue early types amount to 10\% of early types, which 
affects the morphology fraction as a function of
luminosity, the type-specific luminosity functions, and so on.
In this manner, the color gradient and concentration constraints in our
morphological classification reduce the misclassification
of red late type as early type, and blue early type as late type.
This is one of the major improvements we made in this study compared
with previous works.
% -- [end]
Late type galaxies brighter than $M_\ast$ are redder and have more
negative color gradients than fainter late types.
\begin{figure*}
\plotone{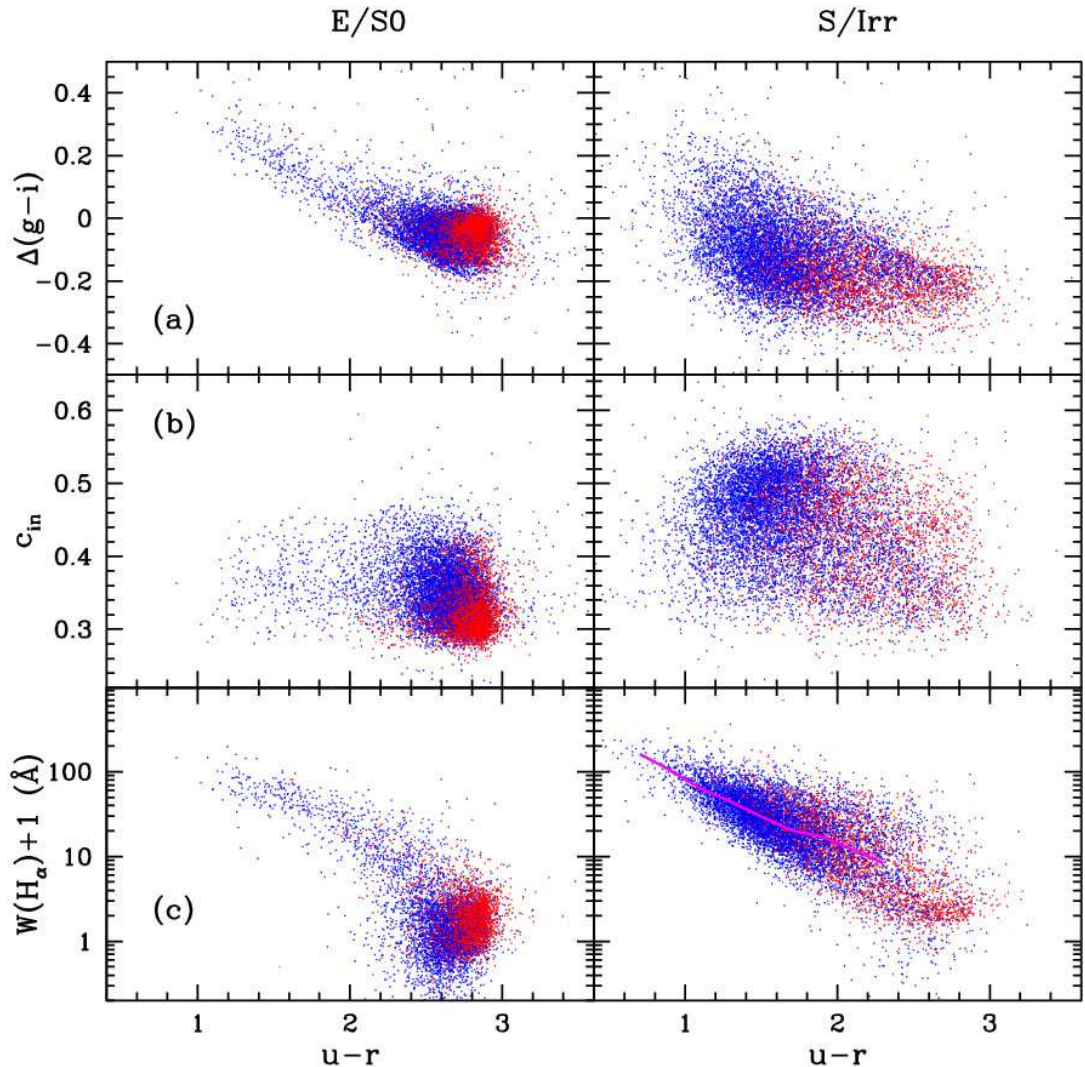}
\caption{
Relations between the $u-r$ color and ($a$) $g-i$ color gradient, ($b$)
(inverse) concentration index, and ($c$) the equivalent width of 
the $H{\alpha}$ line. The most likely $H{\alpha}$ equivalent width
is drawn for the late types ({\it lower right}). Blue points are 
fainter galaxies with $M_r \ge -20.0$, and red points are those
with $M_r < -20.0$.
}
\label{fig4}
\end{figure*}
Figure 6$b$ shows galaxies in the $u-r$ versus the (inverse)
concentration index space. The brightest early types have $u-r \approx
2.9$ and $c_{\rm in}\approx 0.3$.  There is a weak tendency for
the redder ($2.6 \leq u-r \leq 3.0$) bright early types
to have smaller $c_{\rm in}$ than the bluer ($2.3 \leq u-r \leq 2.6$) 
fainter ones.
Late types are loosely concentrated at $u-r\approx 1.5$ and 
$c_{\rm in}=0.49$.  They have a broad overlap with the region 
occupied by early type galaxies.

Figure 6$c$ shows the relations between $u-r$ color and the equivalent
width of $H{\alpha}$ line, $W(H{\alpha})$.  Most red early type
galaxies have weak emission, $W(H{\alpha})<3$. Because galaxies in
the blue trail of early types usually have blue star-forming centers,
$W(H{\alpha})$ in the light from their centers has a tight
correlation with their color. The right panel of Figure 6c shows
that the $H{\alpha}$ emission line strength of the centers of late
type galaxies (plotted as $\log[W(H\alpha)]$) is close to being
linearly proportional to the overall $u-r$ color. This is particularly
true for blue late types.  This proportionality is as expected if
$H{\alpha}$ emission is proportional to $u$-band
luminosity. 
It is interesting to observe that the sequence formed by the blue
early types is far from a straight line and 
has higher $W(H{\alpha})$ than the sequence of the blue late types.

\subsection{Velocity Dispersion of Early Types}

Stellar velocity dispersion is a dynamical indicator of the mass of galaxies.
Figure 7$a$ shows that there is a good correlation 
between velocity dispersion and concentration index, and thus
between mass and surface brightness profile.
Here it should be noted that the D2 sample 
starts to be incomplete at $\sigma\le 150$ km s$^{-1}$ due to the
absolute magnitude cut and the significant dispersion of $\sigma$ at a
given absolute magnitude (see Fig. 4$b$).

The relation becomes tighter for high-$\sigma$ early type
galaxies and approaches the de Vaucouleurs profile value of 
$c_{\rm{in}} = 0.29$ at the highest end.
The size of galaxies is well correlated with velocity
dispersion when $\sigma > 100$ km s$^{-1}$, but the dependence become weaker 
at lower $\sigma$ (Figure 7b).
%This is rather surprising because size and velocity dispersion
%of all galaxies in D2 show a very good correlation with absolute
%magnitude (see Figure 4a and e).
%Down to the luminosity we explore ($M_{r} \sim -18$), there
%seems to be a minimum galaxy size of $1 \sim 2$ h$^{-1}$kpc.
Note that the number of galaxies with $R_{\rm Pet} \leq 3 h^{-1}$kpc
is underestimated in the D2 sample.
The error in $\sigma$ is typically 11 km s$^{-1}$ (see below), which somewhat 
broadens the size versus $\sigma$ relation. 
These systematic effects should be taken into account in interpreting
Figure 7$b$.

Figure 7$c$ indicates that there are progressively more centrally
star-forming early type galaxies as the velocity dispersion
decreases, consistent with the relation between $W(H{\alpha})$
and $M_r$ (Fig. 4$c$).
\begin{figure}
%\epsscale{1.1}
\plotone{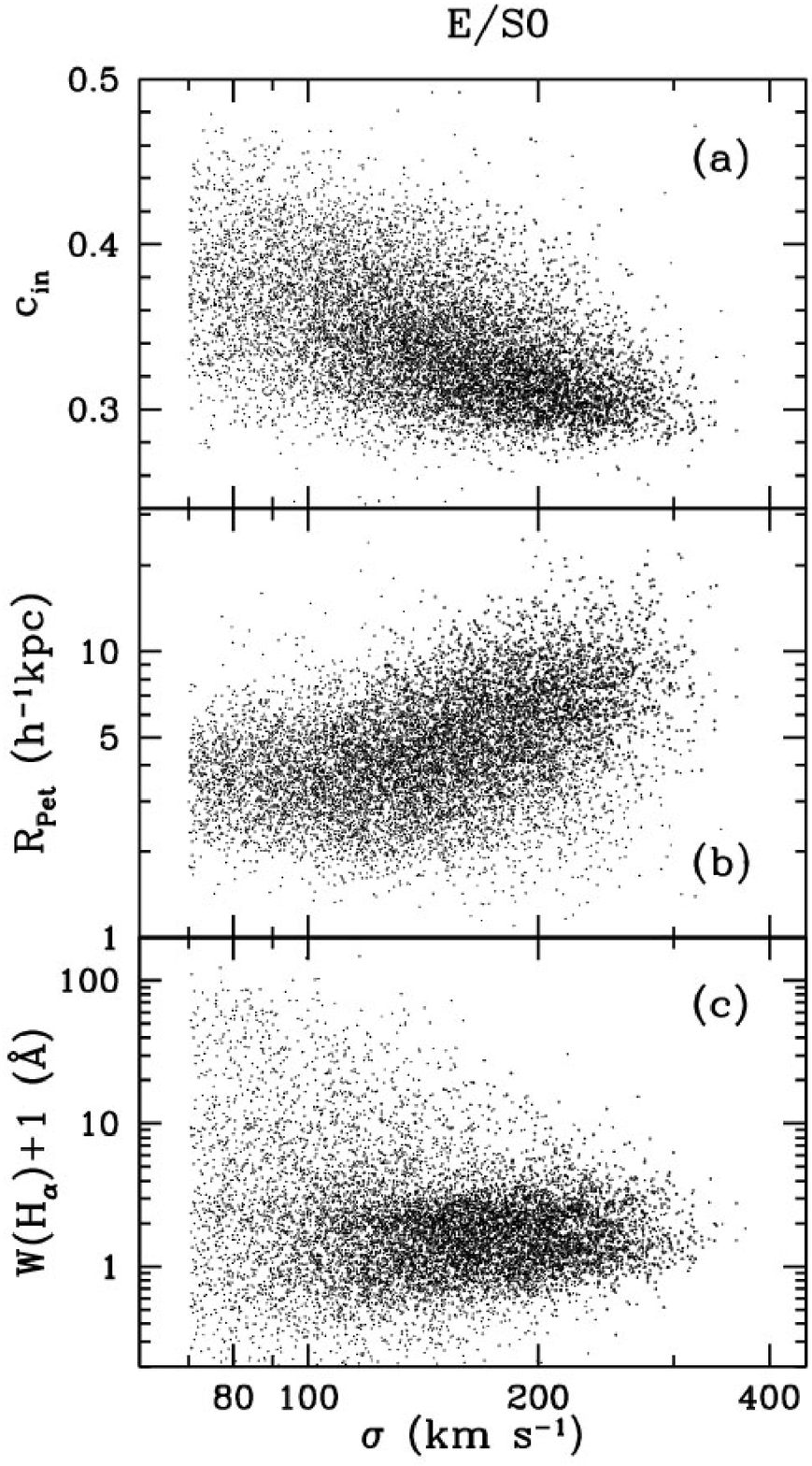}
\caption{Relation between the velocity dispersion of early type galaxies
in sample D2 and ($a$) (inverse) concentration  index, ($b$) 
the Petrosian radius, and ($c$) 
the equivalent width of the $H{\alpha}$ line.
Larger dots are galaxies with $M_{r} < -20.5$, and smaller dots
are those with $M_{r} > -20.5$.
}
\label{fig5}
\end{figure}

\section{Collective Physical Parameters}
\subsection{Luminosity Function}

The upper panel of Figure 8 shows the luminosity function (LF) of
galaxies in the D3 sample. Filled circles are the LF measured by using
both early and late type galaxies. Open circles are the LF of the E/S0 morphological types,
and open squares are that of the S/Irr types. Symbols are calculated
by binning galaxies in each magnitude bin.  Uncertainty limits are
estimated from 16 subsets of each sample.  Curves are the best-fit
Schechter functions of the following form
\begin{equation}
\phi(L)dL = \phi{_*} (L/L{_*})^{\alpha} {\rm exp}(-L/L{_\ast}) dL/L{_\ast}.
\end{equation}
We use the MINUIT package of the CERN program library\footnote{
\url{http://http://wwwasdoc.web.cern.ch/wwwasdoc/minuit/minmain.html}}
to determine the parameters in the Schechter 
function by the maximum likelihood method,
which is applied to individual galaxies 
as described in Sandage et al. (1979).
The Schechter function fits the data extremely well except for the 
highest luminosity bins. 
To reduce the bias in LF due to the internal absorption,
we have included only those late types with axis ratios $b/a>0.6$
in our analyses. To estimate the number density (or $\phi_\ast$) of
all spiral galaxies, we need to know the fraction of late type galaxies
with apparent axis ratios greater than $0.6$.
We adopt the fraction $0.505$ calculated from the CM sample 
(see Fig. 13$b$, below),
whose spiral membership is considered to be the least affected by
the internal absorption. We use this fraction to scale up the late 
type galaxy LF
estimated by using only those with $b/a>0.6$.
The results of the Schechter function fits are summarized in Table 2.
%%%%%%%  TABLE 2 %%%%%%%%%%
\begin{deluxetable*}{lcccc}
\tabletypesize{\small}
\tablecolumns{4}
\tablewidth{0pt}
\tablecaption{Total and Type-specific Schechter Function Parameters}
\tablehead{
\colhead{Name} &\colhead{Absolute Magnitude}
               &\colhead{$M_{\ast}-5{\rm log_{10}}{h}$} 
               &\colhead{$\alpha$}
               &\colhead{$\phi_{\ast}$}\\
            &&&&\colhead{$10^{-2}h^{3}{\rm Mpc}^{-3}$}
}
\startdata
\sidehead{All Types:}
D1&$-18.0>M_{\rm r}$& $  -20.31\pm 0.04$ &$-0.918\pm 0.027$ &$ 1.64$  \\
D2&$-18.5>M_{\rm r}$& $  -20.22\pm 0.03$ &$-0.843\pm 0.026$ &$ 1.72$  \\
D3&$-19.0>M_{\rm r}$& $  -20.22\pm 0.03$ &$-0.807\pm 0.028$ &$ 1.85$  \\
D4&$-19.5>M_{\rm r}$& $  -20.28\pm 0.02$ &$-0.867\pm 0.030$ &$ 2.01$  \\
D5&$-20.0>M_{\rm r}$& $  -20.32\pm 0.02$ &$-0.895\pm 0.039$ &$ 1.87$  \\
\sidehead{Early Types:}
D1&$-18.0>M_{\rm r}$& $  -20.17\pm 0.05$ &$-0.494\pm 0.047$ &$ 0.73$  \\
D2&$-18.5>M_{\rm r}$& $  -20.16\pm 0.04$ &$-0.473\pm 0.044$ &$ 0.71$  \\
D3&$-19.0>M_{\rm r}$& $  -20.23\pm 0.04$ &$-0.527\pm 0.043$ &$ 0.71$  \\
D4&$-19.5>M_{\rm r}$& $  -20.38\pm 0.03$ &$-0.719\pm 0.042$ &$ 0.75$  \\
D5&$-20.0>M_{\rm r}$& $  -20.49\pm 0.03$ &$-0.870\pm 0.052$ &$ 0.66$  \\
\sidehead{Late Types:}
D1&$-18.0>M_{\rm r}$& $  -20.30\pm 0.05$ &$-1.070\pm 0.033$ &$  0.98$  \\
D2&$-18.5>M_{\rm r}$& $  -20.18\pm 0.04$ &$-0.965\pm 0.034$ &$  1.10$  \\
D3&$-19.0>M_{\rm r}$& $  -20.12\pm 0.03$ &$-0.897\pm 0.037$ &$  1.23$  \\
D4&$-19.5>M_{\rm r}$& $  -20.09\pm 0.03$ &$-0.832\pm 0.043$ &$  1.44$  \\
D5&$-20.0>M_{\rm r}$& $  -20.04\pm 0.03$ &$-0.666\pm 0.061$ &$  1.37$  \\
\enddata
%\tablecomments{}
\end{deluxetable*}
%%%%%%%%%%%%%%%%%%%%%%%%%%%%%%%%%%%

Since our volume-limited samples from the SDSS do not extend
to very faint magnitudes, there is a significant correlation between
measured $M_\ast$ and $\alpha$. 
We choose the D3 sample to estimate a representative LF of the SDSS
galaxies as a compromise; 
the uncertainty in $M_*$ is minimized for D4, while the uncertainty in
$\alpha$ is minimized for D2. However, the D3 sample covers the absolute
magnitude space brighter than $-19.0$, only about 1.2 mag fainter than
$M_\ast$, and the correlation between the estimated parameters $M_\ast$
and $\alpha$ makes their physical meanings less clear. 
The LF of the deeper D4 sample, which includes the
Sloan Great Wall (Gott et al. 2005), shows the highest density of
early type galaxies among all five volume-limited samples, and thus
its LF seems less representative. The D5 sample does not extend faint
enough to determine the $\alpha$ parameter.  
%--- [This part is added]
The existence of the large scale structure affects both normalization
and shape of LF (e.g. Croton et al. 2005).
Therefore, the trends in $M_\ast$ and $\alpha$ are affected by
the fact that nearby universe is a relatively underdense region and 
the shallower samples are more dominated by the relatively faint galaxies
which prefers underdense regions (see Fig. 10 of Paper II for further
details).
%--[end]
When all spiral galaxies are used (i.e. no internal absorption effect
correction),
the LF of all galaxies in the D3 sample is well fitted by $M_\ast =
-20.24\pm0.03$ and $\alpha=-0.90\pm0.03$.
For comparison, Blanton et al. (2003a) report
$M_\ast =-20.44\pm 0.01$, and $\alpha=-1.05\pm0.01$ for 
an apparent magnitude-limited sample called
SDSS LSS Sample 10 (similar to SDSS DR1).  If we adopt the faint
end slope $\alpha$ from the shallow sample D1 and $M_\ast$ from the
deepest sample D5, our estimates are $\alpha=-0.99\pm0.03$ and
$M_\ast =-20.38\pm0.03$, which are very close to Blanton et al.'s.  
%The systematic changes in $M_{*}$ and $\alpha$ of different volume-limited
%samples are caused by the luminosity-local density relation (Park et
%al. 2006).

When the internal absorption effect in the sample is reduced,
the LF of spirals changes, particularly in the $\alpha$ parameter.
The LF of early types in sample D3 has $M{_\ast}=-20.23\pm 0.04$ and
$\alpha = -0.53\pm 0.04$ while that of late types has a fainter
characteristic magnitude of $M{_*}=-20.12\pm 0.03$ and a steeper
faint-end slope of $\alpha = -0.90\pm 0.04$.  
%-- [this part is added]
If all spiral galaxies are used, the LF of late types
has $M{_*}=-20.09\pm 0.04$ and $\alpha = -1.01\pm 0.04$.
%-- [end]
This demonstrates that the faint-end slope $\alpha$ of late types would have been
misleadingly measured steeper if the inclined late type galaxies
had not been excluded.
The blue galaxies
classified as early types by our morphological classifier make an
increasing contribution to the LF at faint magnitudes.  We suspect
that some of these objects are actually bulge-dominated spirals whose disks
are lost to the background sky.  However, the absolute contribution of
blue early type galaxies to the LF is still small at the magnitudes
studied, as seen in Figure 3$a$, and the type-specific LFs at
magnitudes brighter than $M_r = -18$ seems well determined by
our samples.
Baldry et al. (2004) used $u-r$ color-selected early (red)
and late (blue) type galaxy samples to measure their luminosity
functions.
The estimated $M_\ast$ parameter differed by 0.21 mag
and the $\alpha$-parameter was measured to be $-0.83$ (early)
and $-1.18$ (late). The difference in $M_\ast$
between early and late types is quite consistent with our result,
but they obtained steeper $\alpha$'s.

The bottom panel of Figure 8 shows the fraction of E/S0 galaxies in
all five volume-limited samples as a function of $r$-band absolute
magnitude.  Again, only late types with $b/a>0.6$ are used, and their
number fraction of 0.505 is used to infer the total number of spiral
galaxies.  There is a surprisingly consistent monotonic relation
between luminosity and early type galaxy fraction. The critical
magnitude at which the early type fraction exceeds 50\% is $M_r =
-21.29\pm 0.04$.  Even though the shape of LF itself shows
significant fluctuations due to large scale structures when measured in
different regions of the universe, the morphological fraction as a
function of luminosity is relatively less sensitive to large scale structures
and thus seems to be more universal.
However, it is well-known that the early type fraction is 
a monotonically increasing function of local density
(Dressler 1980; Tanaka et al. 2005; Goto et al. 2003a; Paper II).
Weinmann et al. (2006) also measured the morphology fraction
as a function of $M_r$ and found basically the same dependence,
even though the sample they used is smaller.
\begin{figure}
\epsscale{1.2}
\plotone{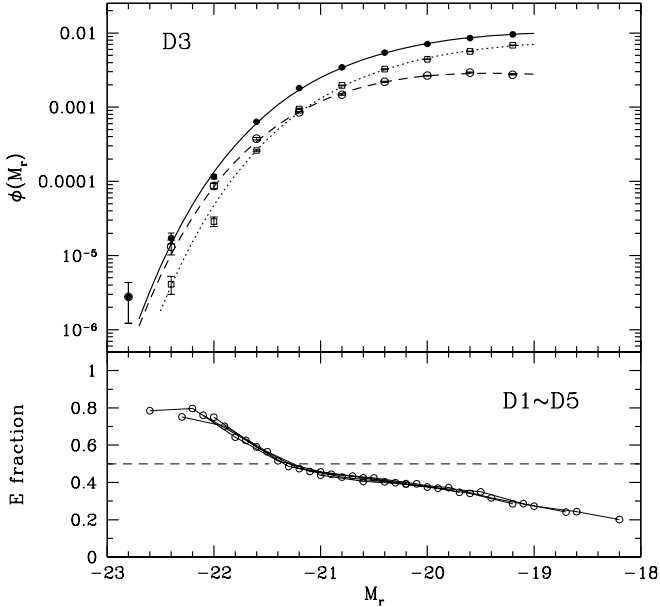}
\caption{
{\it Top}: Luminosity function (LF) of galaxies in the D3 sample. 
Filled circles are for all galaxies, open circles for the
E/S0 morphological types, 
and open squares for the S/Irr types. Symbols are the LF calculated 
by binning galaxies in each magnitude bin. Curves are the best-fit
Schechter functions.
{\it Bottom}: The elliptical galaxy fraction in magnitude bins measured from
the five volume-limited samples (D1 - D5).
}
\label{fig6}
\end{figure}

\subsection{Velocity Dispersion Function}
The velocity dispersion function (VDF) of early type galaxies can be
used to derive gravitational lens statistics and to estimate
cosmological parameters.  To estimate the velocity dispersion function
we need to have a sample of galaxy velocity dispersion data
that is complete over a wide range. Figure 4$b$ indicates that sample D2, for
example, starts to be incomplete at $\sigma\le 150$ km s$^{-1}$ due to the
absolute magnitude cut and the significant dispersion of $\sigma$ at a
given absolute magnitude. To lower the completeness limit down to
$\sigma\approx 70$ km s$^{-1}$ we have used a series of our volume-limited
samples an additional very faint volume-limited sample with 
$-16.8\ge M_r > -17.5$, which is added because even the CM sample is not faint
enough to give the $\sigma$-distribution complete down to
$\sigma=70$ km s$^{-1}$. The differences in volumes of these volume-limited
samples are taken into account when we estimate the velocity
dispersion function from a combination of seven samples.

The circles in Figure 9 
show the VDF of early type galaxies measured in this way.
\begin{figure}
\epsscale{1.2}
\plotone{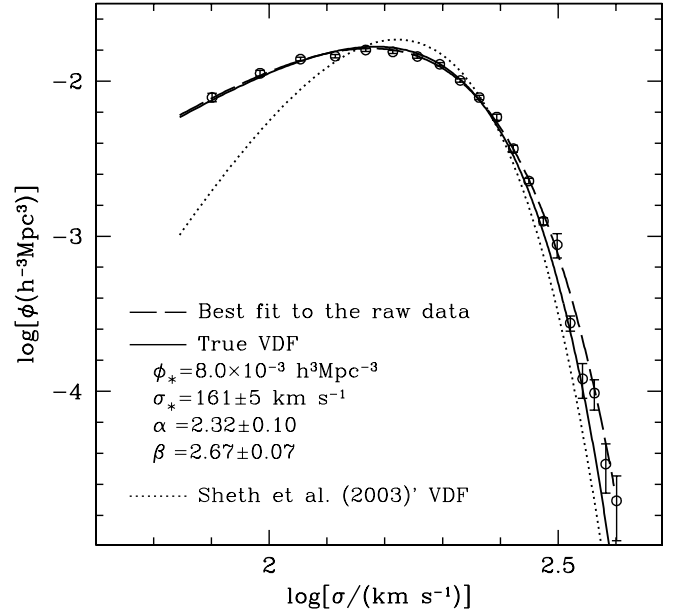}
\caption{Velocity dispersion functions (VDFs; {\it circles}) 
measured from our samples with galaxies brighter than $M_r = -16.8$.
The solid curve is our estimate of the true VDF best fit 
by a modified Schechter function. 
The long-dashed curve is the best fit to the raw measurement.
The dotted curve is the fit given by Sheth et al. (2003).}
\end{figure}
\begin{figure}
\epsscale{1.2}
\plotone{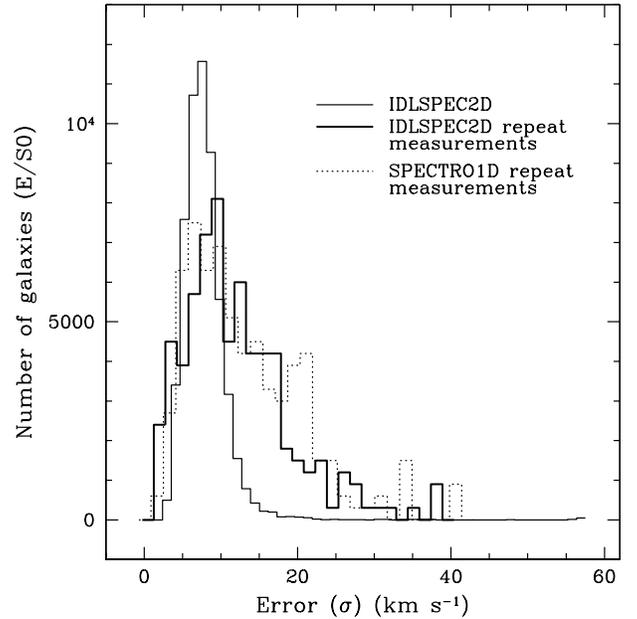}
\caption{Histogram showing the distributions of the standard deviations
of $\sigma$ of the galaxies measured by two different SDSS pipelines
{\tt IDLSPEC2D} ({\it thin and thick solid lines}) and {\tt SPECTRO1D}
({\it dotted line}). The thick solid and dotted lines are obtained from
the repeat measurements (plate 406 and 412) and the thin solid line
is from the formal error given by the {\tt IDLSPEC2D}.
}
\end{figure}
The curve is the fit by a modified Schechter function of the form
\begin{equation}
\phi(\sigma)d\sigma = \phi{_\ast} (\sigma/\sigma{_\ast})^{\alpha} 
                      {\rm exp}[-(\sigma/\sigma{_\ast})^\beta]{\beta \over 
                      \Sigma(\alpha/\beta)}{d\sigma \over \sigma},
\end{equation}
where $\sigma{_\ast}$ is the characteristic velocity dispersion, $\alpha$ is 
the low-velocity power-low index, and $\beta$ is the high-velocity
exponential cutoff index (Mitchell et al. 2005).
The VDF parameters obtained by the MINUIT package are
\begin{eqnarray*}
(\phi_\ast, \sigma_\ast, \alpha, \beta) &=&
(8.0\times 10^{-3} h^3 {\rm Mpc}^{-3},
~~161 \pm 5 {\rm km s^{-1}}, \\
&&2.32\pm 0.10,~~ 2.67\pm 0.07).
\end{eqnarray*}
When we made the fit with the above parameters, we assumed that the measurement
error in $\sigma$ was approximated by $0.084\sigma$ km s$^{-1}$ (see below).
The dashed curve in Figure 9 is the modified Schechter function best fit to
the raw data points ({\it open circles}) which are broadened at high 
$\sigma$'s due to the measurement error. 
The solid curve is the deconvolved VDF whose parameters are given above.

It should be noted that there exists very strong correlations
among the fitting parameters, and the results of the fitting
depend sensitively on the range of the velocity dispersion used.
Our estimate of the VDF of early type galaxies is quite different 
from that of Sheth et al. (2003) and Mitchell et al. (2005).
The dotted curve in Figure 9 is the VDF best fit given by
Sheth et al. (2003).
Mitchell et al. (2005) presented the same function with the 
normalization $\phi_\ast$ lowered by 30\%.
The difference at small $\sigma$ seems due to the fact that
some of morphologically early type galaxies are discarded
based on spectrum in Sheth et al. (cf. Fig. 12 of Mitchell et al. 2005).
It is not clear what makes the difference at large $\sigma$,
but it could be due to the measurement error larger than that adopted here.

In order to check whether or not our VDF is seriously affected 
by the measurement error in $\sigma$ we have made the following test. 
We first adopt Sheth et al. (2003)'s VDF as the true one. 
We then look for galaxies 
having 4 or 5 repeat measurements of $\sigma$ in southern stripe
82 (plates 406 and 412) to estimate the error.  The thick histogram
in Figure 10 shows the distribution of the standard deviations 
of $\sigma$ of the galaxies with repeat measurements. 
The median size of error is 11 km s$^{-1}$.  One can note
that the error estimated from the repeat measurements by the {\tt IDLSPEC2D}
is larger than the formal error ({\it thin histogram}) given by the {\tt IDLSPEC2D}, 
but is comparable with that ({\it dotted histogram}) 
obtained from the repeat measurements 
by the official spectroscopic pipeline, {\tt SPECTRO1D} (written by 
M. SubbaRao, M. Bernardi and J. Frieman). 
If the measurement error were larger than 20 km s$^{-1}$, the dispersion
of the relation between $M_r$ and $\sigma$ shown in Figure 4$b$
would be much larger.
We then calculate the VDF affected by the error in $\sigma$.
\begin{figure}
\epsscale{1.2}
\plotone{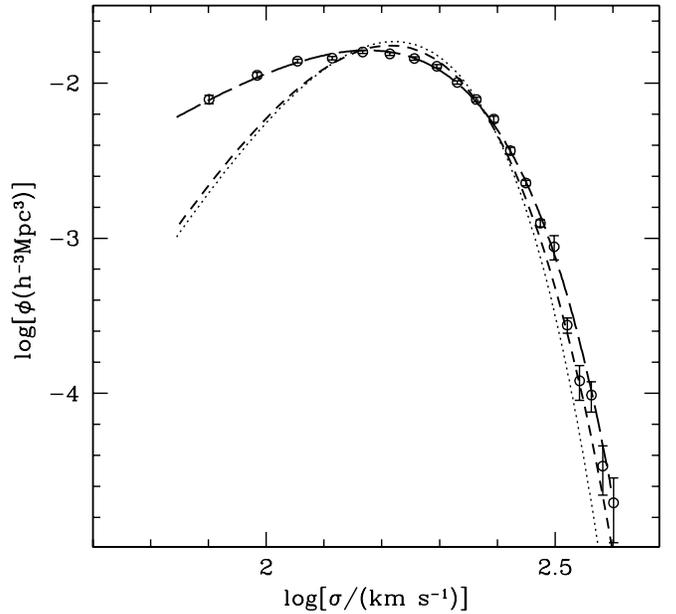}
\caption{Velocity dispersion functions (VDFs) demonstrating the effect
of the measurement errors in $\sigma$ on the VDF. 
The dotted curve is Sheth et al.(2003)'s VDF, and the short-dashed curve is
the VDF deformed due to the Gaussian-distributed measurement error of
$0.084\sigma$ km s$^{-1}$. The long-dashed curve is the best fit to
the VDF measured in this work.}
\end{figure}
The dotted curve in Figure 11 is Sheth et al.'s function, 
and the short-dashed curve is the 
VDF deformed due to the Gaussian-distributed measurement error of 
$0.084\sigma$ km s$^{-1}$, 
which reflects the fact that the error in $\sigma$ estimated 
from the repeat measurements increases as $\sigma$ increases. 
It demonstrates that our VDF cannot be realized from Sheth et al.'s
function by the convolution effects of the measurement error

\subsection{Axis Ratio Distribution}

The scatter plot of galaxies in Figure 5 indicates
that the mean axis ratio of bright late type galaxies 
tend to be high. This is due to the internal absorption making inclined 
galaxies appear fainter. The bottom panel of Figure 12 shows the late type 
galaxies of sample D3 in the $M_r$ versus $u-r$ color space as in 
the right panel of Figure 3$a$. Galaxies with $b/a > 0.8$ are shown 
as blue dots, and those with $b/a<0.4$ are shown as red dots. 
Their most likely color curves are also drawn. We find that the group
of highly inclined galaxies is significantly shifted with respect to the almost
face-on ones toward fainter magnitudes and redder colors. We also find
that the shapes of the two distributions are different, which can be
explained if the galaxies with intermediate magnitude ($M_r\sim
-20.5$) and color ($u-r\sim 1.6$) suffer from more dimming and
reddening due to internal extinction. The color of very red or very
blue galaxies seems to be less affected by the internal extinction. A
detailed study of internal extinction is needed for the SDSS galaxies.
%-----
The early types are not much affected.
%--- [end]
If one does not take into account this inclination effect on
magnitude, the measured LF of late type galaxies and the morphology
fraction will be seriously in error. In analyses where these
systematics are important, we have used late types with $b/a
>0.6$. Figure 5 demonstrates that the inclination effects on
magnitude is small for this set of late type galaxies.

\begin{figure*}
\epsscale{0.9}
\plotone{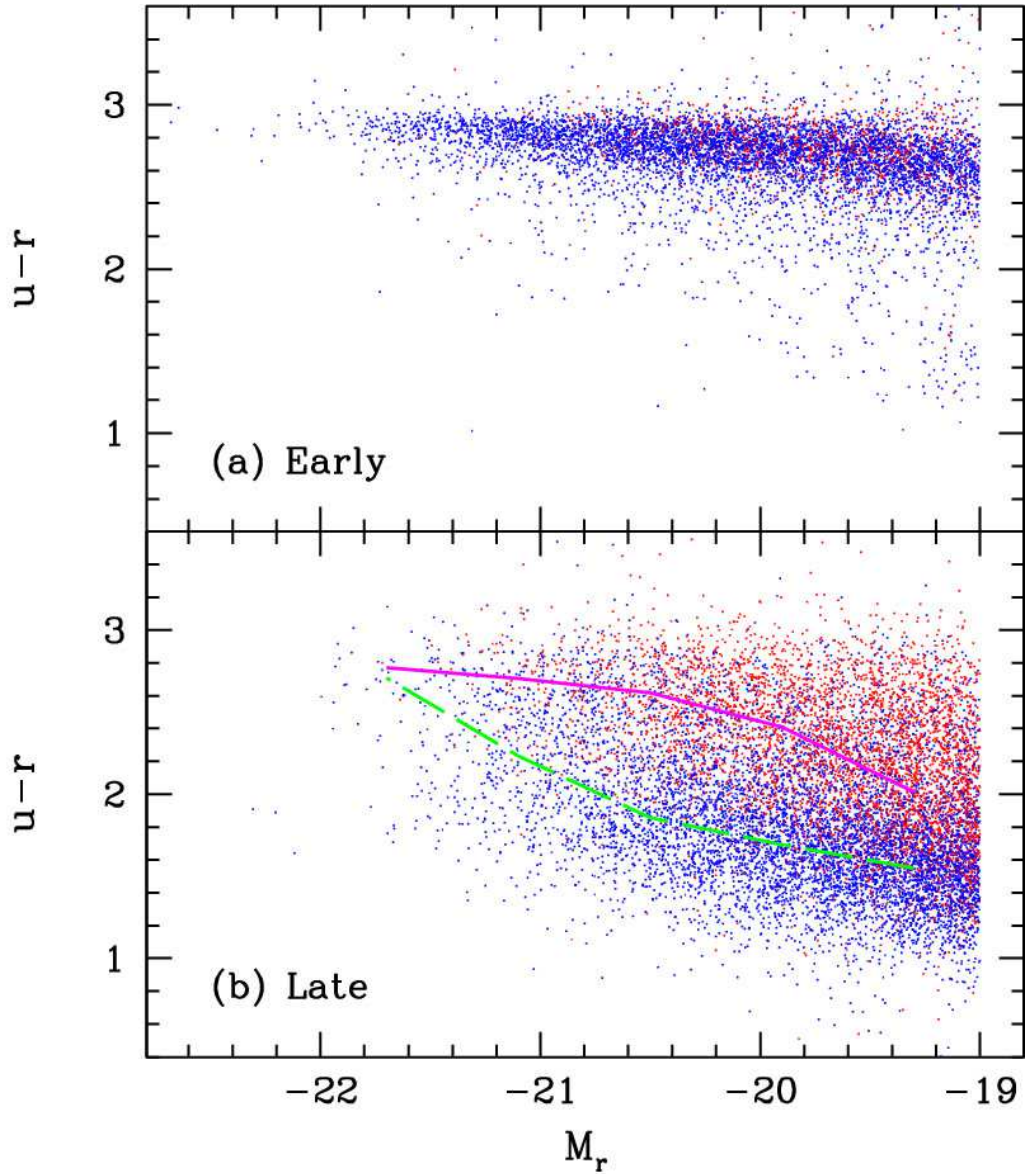}
\caption{Distributions of ($a$) early and ($b$) late type galaxies in the
color magnitude diagram.
Galaxies with $b/a > 0.8$ are shown as blue dots,
and those with $b/a < 0.4$ are shown as red dots.
For late types the most likely color-magnitude relations
are drawn for the different axis ratio sets.
}
\label{fig8}
\end{figure*}

\begin{figure}
%\epsscale{0.8}
\plotone{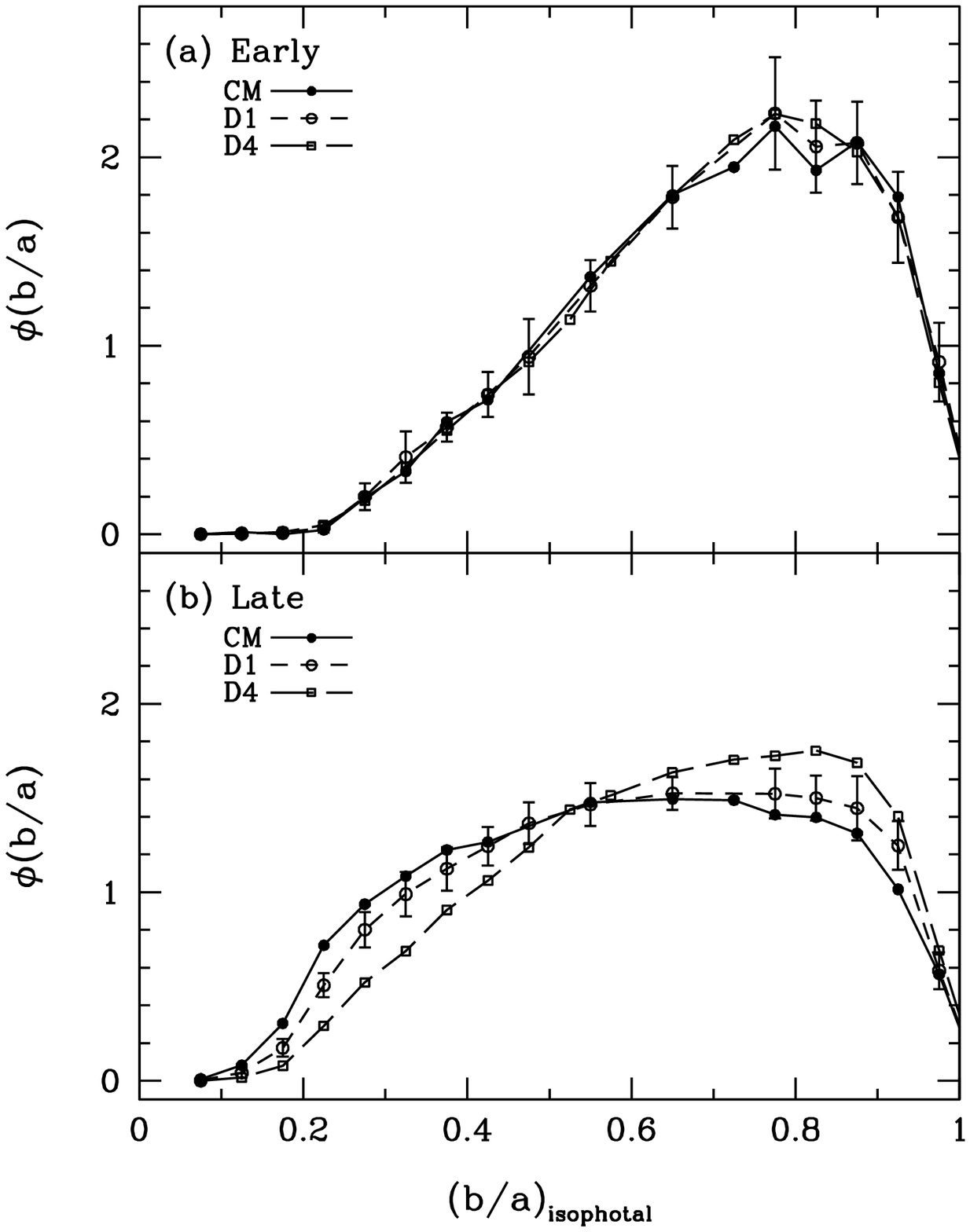}
\caption{Distributions of the apparent isophotal axis ratio of ($a$)
early and ($b$) late type galaxies.
}
\label{fig8}
\end{figure}

Because of the inclination effects on magnitude, our volume-limited
samples are not complete for late type galaxies near the absolute
magnitude cuts. The lower panel of Figure 13 shows distributions of 
isophotal axis ratio of late type galaxies in samples CM 
({\it solid line with filled circles}), D2
({\it short-dashed line with open circles}), and D4 
({\it long-dashed line with open squares}). 
There appears to be no late type galaxy with nearly zero axis
ratio because of the intrinsic thickness of disks of late type
galaxies.  The axis ratio of 0.2 due to the intrinsic thickness 
is consistent with previous works (Sheth et al. 2003).
Almost face-on galaxies are also lacking, because disks of
late type galaxies do not appear exactly round even when they are
face-on (Kuehn \& Ryden 2005).  The
distributions are significantly different from one another because of
the internal absorption effects, lacking highly inclined galaxies in the
samples with bright cuts. We assume that the magnitude limit of
CM sample ($M_{r, {\rm cut}}=-17.5$) is faint enough that its $b/a$
distribution is close to the intrinsic distribution.
The fraction of galaxies with $b/a>0.6$
in the CM sample is 0.505.  The upper panel of Figure 13 shows the
distribution of projected axis ratio of early type
galaxies. Distributions obtained from three samples with different
magnitude cuts match one another very well.  This reflects the
fact that internal absorption in early type galaxies is negligible.

Alam \& Ryden (2002) measured the axis ratio distributions for
red and blue populations in the SDSS EDR sample.
The distribution functions for red or blue galaxies well-fitted by
a de Vaucouleurs profile match well with our result for early types.
Sheth et al. (2003) also measured the distribution of
axis ratio of early and late type galaxies in the SDSS.
Their axis ratio distribution of early types agrees
well with ours except at the $b/a = 1$ bin.

\section{Summary}

We have elaborately investigated the relations among various
internal (e.g. color, luminosity, morphology, star formation rate, 
velocity dispersion, size, radial color gradient, and axis ratio) 
and collective properties (e.g. luminosity function, velocity
dispersion distribution function, and  axis ratio distribution function).
We will study other collective properties such as the correlation
function, power spectrum, topology, and peculiar velocity field,
in separate works.

There are a number of important improvements we made 
in our measurement:
\begin{enumerate}
\item To extend the absolute magnitude range,
we have added the redshifts of the bright galaxies with $r<14.5$
to the SDSS-NYU-VAGC. Six volume-limited samples with
faint limits from $M_r = -17.5$ to $-20.0$, are
used for more efficient use of the flux-limited SDSS sample.
\item We have divided the samples into
subsamples of early (E/S0) and late (S/Irr) morphology types
by using our accurate morphology classifier, working in 
the three-dimensional parameter space of
$u-r$ color, $g-i$ color gradient, and concentration index.
The morphology classifier is able to separate the blue early
types from the blue late types, and also to distinguish
the spirals from the red E/S0 types because it does not
depend only on color.
\item The late type galaxies with isophotal
axis ratio of less than 0.6 have been excluded
because they suffer severely from
dimming and reddening due to internal extinction
and then may cause biases in luminosity, luminosity function,
color, color gradient, and so on.
This correction makes the blue sequence of the late type galaxies
bluer and brighter compared to that obtained from all
late type galaxies. The blue sequence also become better-defined
with smaller width.

\end{enumerate}

We have found that 
absolute magnitude and morphology are the most important
parameters in characterizing physical properties of galaxies.
In other words, other parameters show relatively small 
scatter once absolute magnitude and morphology are fixed
for early type galaxies, in particular.
One very interesting fact we have noticed in our work
is that many physical parameters of galaxies manifest 
different behaviors across the absolute magnitude of
about $M_\ast \pm 1$.
For example, the red sequence of early types changes the slope
at $M_{\rm r} \approx -19.6$ in the color-magnitude
space. Also, at a magnitude fainter than $M_{\rm r} \sim -20$,
the number of blue star-forming early types
increases significantly, and the surface brightness profile
as parameterized by $c_{\rm in}$ becomes less centrally concentrated.
The passive spirals with vanishing $H_\alpha$ emission
seem to have magnitudes brighter than $M_{\rm r} \sim -19.5$.

At fixed morphology and luminosity, we find that bright ($M_r<-20$)
early type galaxies show very small dispersions in color, color
gradient, concentration, size, and velocity dispersion. These
dispersions increase at fainter magnitudes, where the fraction of 
blue star-forming early types increases. 
Concentration indices of early types are well-correlated with 
velocity dispersion, but are insensitive to luminosity and color 
for bright galaxies, in particular. The slope of the Faber-Jackson 
relation ($L \propto {\sigma}^{\gamma}$) continuously changes 
from $\gamma=4.6\pm 0.4$ to $2.7\pm 0.2$ when luminosity 
changes from $M_r = -22$ to $-20$.  
The size of early types is well-correlated with stellar velocity
dispersion, $\sigma$, when $\sigma>100$ km s$^{-1}$.
Late type galaxies show 
wider dispersions in all physical parameters compared to early 
types at the same luminosity.
We find that passive spiral galaxies are well-separated from
star-forming late type galaxies at $H{\alpha}$ equivalent width of
about 4.

We note that the estimated LF shows significant fluctuations 
due to large scale structures when measured in
different regions of the universe.
On the other hand, the morphological fraction as a
function of luminosity is relatively less sensitive to large scale structures
and thus seems to be more universal. 
In Paper II, it is shown that the morphology
fraction is a monotonic function of local density at a fixed luminosity.
Since luminosity in turn depends on local density,
this result indicates that
the probability for a galaxy to be born as a particular morphological
type is basically determined when its luminosity or mass
is given. To this extent, morphology is determined by 
galaxy mass. The question here is
what determines the morphology of a galaxy 
having a particular mass while keeping the type fraction 
corresponding to its mass scale. Further investigations are required 
to answer this question. 

\acknowledgments
The authors thank Chan-Gyung Park for making the velocity dispersion function 
fits and Ravi Sheth for helpful comments.
CBP acknowledges the support of the Korea Science and Engineering
Foundation (KOSEF) through the Astrophysical Research Center for the
Structure and Evolution of the Cosmos (ARCSEC).
MSV acknowledges support from NASA grant
NAG-12243 and NSF grant AST-0507463. MSV thanks the Department of
Astrophysical Sciences at Princeton University for its hospitality
during sabbatical leave. YYC, CBP, and MSV thank the Aspen Center for
Physics, at which much of this paper was written.

Funding for the SDSS and SDSS-II has been provided by the Alfred P. 
Sloan Foundation, the Participating Institutions, the National Science 
Foundation, the U.S. Department of Energy, the National Aeronautics 
and Space Administration, the Japanese Monbukagakusho, the Max 
Planck Society, and the Higher Education Funding Council for England. 
The SDSS Web Site is http://www.sdss.org/.

The SDSS is managed by the Astrophysical Research Consortium for the 
Participating Institutions. The Participating Institutions are the 
American Museum of Natural History, Astrophysical Institute Potsdam, 
University of Basel, Cambridge University, Case Western Reserve 
University, University of Chicago, Drexel University, Fermilab, 
the Institute for Advanced Study, the Japan Participation Group, 
Johns Hopkins University, the Joint Institute for Nuclear Astrophysics, 
the Kavli Institute for Particle Astrophysics and Cosmology, 
the Korean Scientist Group, the Chinese Academy of Sciences (LAMOST), 
Los Alamos National Laboratory, the Max-Planck-Institute for Astronomy, 
the Max-Planck-Institute for Astrophysics, New Mexico State University, 
Ohio State University, University of Pittsburgh, University of 
Portsmouth, Princeton University, the United States Naval Observatory, 
and the University of Washington.

\end{document}